\documentclass[sigconf]{acmart}

\AtBeginDocument{%
  }

\setcopyright{acmlicensed}
\copyrightyear{2018}
\acmYear{2018}
\acmDOI{XXXXXXX.XXXXXXX}
\acmConference[Conference acronym 'XX]{Make sure to enter the correct
  conference title from your rights confirmation email}{June 03--05,
  2018}{Woodstock, NY}
\acmISBN{978-1-4503-XXXX-X/2018/06}

\usepackage{microtype}

\usepackage{graphicx}
\usepackage{amsmath}
\usepackage{mathtools}
\usepackage{algorithm}
\usepackage{latexsym}
\usepackage{microtype}
\usepackage{booktabs}
\usepackage{algpseudocode}
\usepackage{multirow}
\usepackage{subfigure}
\usepackage{balance}
\usepackage{color}
\usepackage{xcolor}
\usepackage{enumitem}
\usepackage{xspace}
\usepackage{setspace}
\usepackage{longtable}
\usepackage{pifont}
\usepackage{graphicx}
\usepackage{subcaption}

\usepackage{marginnote}
\newcommand{\pageenlarge}[1]{\marginnote{}\enlargethispage{#1\baselineskip}}

\usepackage{xcolor}
\usepackage{framed}

\definecolor{mygray}{gray}{0.9}

\usepackage[most]{tcolorbox} 
\usepackage{textcomp}
\usepackage{caption}
\usepackage{tcolorbox}
\tcbuselibrary{breakable, skins}

\definecolor{codegreen}{rgb}{0,0.6,0}
\definecolor{codegray}{rgb}{0.5,0.5,0.5}
\definecolor{codepurple}{rgb}{0.58,0,0.82}
\definecolor{backcolour}{rgb}{0.98,0.98,0.98}

\lstdefinestyle{jsonstyle}{
    language=Java,
    backgroundcolor=\color{backcolour},
    commentstyle=\color{codegreen},
    keywordstyle=\color{magenta},
    stringstyle=\color{codepurple},
    basicstyle=\ttfamily\scriptsize,
    breakatwhitespace=false,         
    breaklines=true,
    captionpos=b,                    
    keepspaces=true,                 
    showspaces=false,                
    showstringspaces=false,
    showtabs=false,                  
    tabsize=2,
    frame=single,
    rulecolor=\color{black!20},
    columns=flexible,
    moredelim=[is][\bfseries]{|}{|},
}
\usepackage[most]{tcolorbox}
\usepackage{fontawesome5}
\usepackage{xcolor}

\newcommand{\cmark}{\textcolor{green!60!black}{\faCheck}} 
\newcommand{\xmark}{\textcolor{red!80!black}{\faTimes}}

\newtcolorbox{modelbox}[3]{
    colback=#1,
    boxrule=0.5pt,
    arc=2pt,
    left=1mm, right=1mm, top=1mm, bottom=1mm,
    fontupper=\tiny,
    fonttitle=\bfseries\footnotesize,
    coltitle=black, 
    title={#2 \hfill #3},
    attach title to upper={\par\vspace{1mm}},
}

\usepackage{dblfloatfix}

\begin{document}

\title{MLDocRAG: Multimodal Long-Context Document Retrieval Augmented Generation}


\author{Yongyue Zhang}
\affiliation{%
  \institution{Independent Researcher}
  \country{Singapore}
  }
\email{yongyue002@gmail.com}

\author{Yaxiong Wu}
\affiliation{%
  \institution{Independent Researcher}
  \country{Singapore}
  }
\email{wuyashon@gmail.com}

\begin{abstract}

Understanding multimodal long-context documents that comprise multimodal chunks such as paragraphs, figures, and tables is challenging due to (1) cross-modal heterogeneity to localize relevant information across modalities, (2) cross-page reasoning to aggregate dispersed evidence across pages.
To address these challenges, we are motivated to adopt a query-centric formulation that projects cross-modal and cross-page information into a unified query representation space, with queries acting as abstract semantic surrogates for heterogeneous multimodal content.
In this paper, we propose a Multimodal Long-Context Document Retrieval Augmented Generation (MLDocRAG) framework that leverages a Multimodal Chunk-Query Graph (MCQG) to organize multimodal document content around semantically rich, answerable queries.
MCQG is constructed via a multimodal document expansion process that generates fine-grained queries from heterogeneous document chunks and links them to their corresponding content across modalities and pages.
This graph-based structure enables selective, query-centric retrieval and structured evidence aggregation, thereby enhancing grounding and coherence in multimodal long-context question answering.
Experiments on datasets MMLongBench-Doc and LongDocURL demonstrate that MLDocRAG consistently improves retrieval quality and answer accuracy, demonstrating its effectiveness for multimodal long-context understanding.

\end{abstract}

\maketitle

\pageenlarge{1} \looseness -1
\section{Introduction}~\label{sect:introduction}

Multimodal long-context documents, such as research papers, reports, and books, often span tens to hundreds of pages and contain diverse multimodal components/chunks including text, images, and tables~\cite{van2023document,park2025dochop,tanaka2023slidevqa,islam2023financebench,cho2024m3docrag}. Understanding such lengthy multimodal documents presents two central challenges~\cite{cho2024m3docrag}: (1) \textit{cross-modal heterogeneity}, which requires identifying and localizing relevant information across heterogeneous modalities; and (2) \textit{cross-page reasoning}, which demands integrating evidence scattered across multiple pages to support coherent inference. 
Addressing these challenges necessitates the ability of \textit{multimodal long-context association}—accurately identifying, connecting, and integrating semantically relevant information across modalities and segments.

Large Vision-Language Models (LVLMs) have shown strong cross-modal understanding capabilities of effectively aligning and interpreting multimodal information within localized short contexts~\cite{hu2024mplug,dong2024internlm,yin2024survey}.
Representative examples include GPT-4o~\cite{hurst2024gpt}, Gemini~\cite{team2023gemini}, Qwen2.5-VL~\cite{bai2025qwen2}, and InternVL~\cite{chen2024internvl}, which demonstrate impressive performance on short-range multimodal reasoning benchmarks~\cite{li2024survey,wang2025mcot}.
However, they often struggle to maintain consistent semantic modeling within limited context windows when applied to long-document scenarios~\cite{liu2025comprehensive}. In particular, their performance deteriorates when evidence is sparsely distributed across pages and modalities by overlooking the relevant information, leading to the so-called ``needle in a haystack'' problem~\cite{wang2025multimodal,wang2024needle}. 

\begin{figure*}[t]
  \centering
  \includegraphics[width=0.98\linewidth]{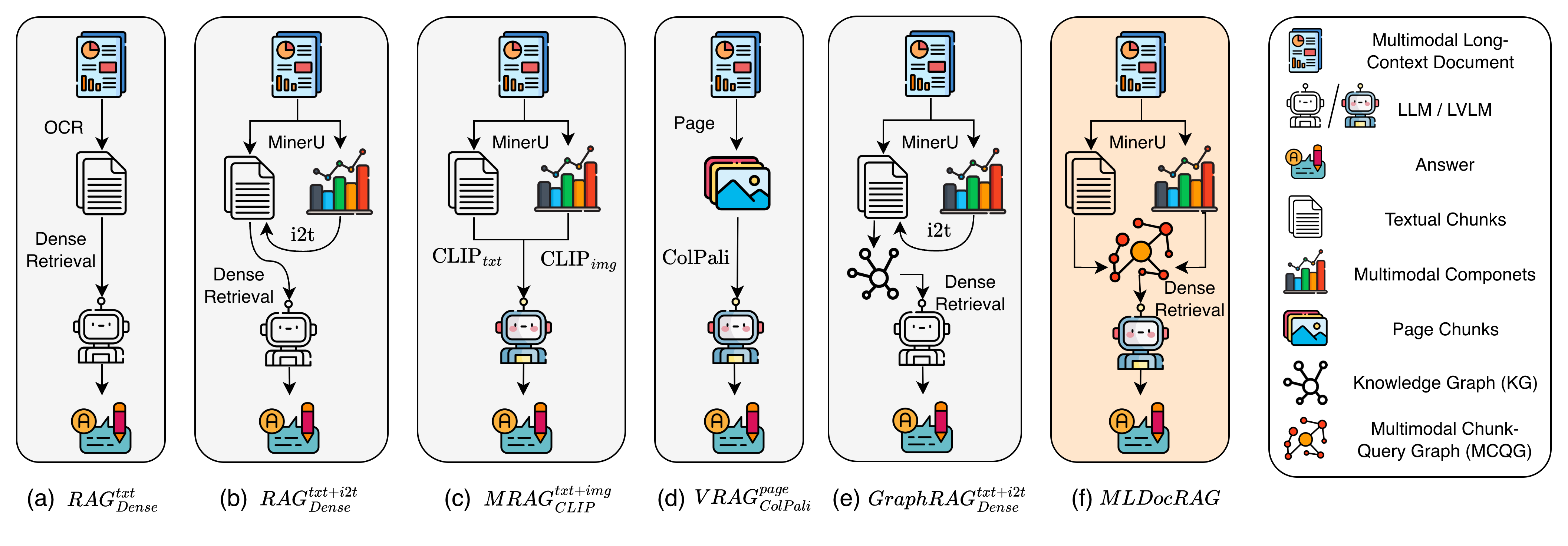}
  \vspace{-1.5\baselineskip}
  \caption{Illustration of RAG for multimodal long-context documents, comparing (a)–(e) baselines with (f) our MLDocRAG.}
  \label{fig:task}
  \vspace{-1.0\baselineskip}
\end{figure*}

\pageenlarge{1} \looseness -1
Retrieval-Augmented Generation (RAG) has emerged as a promising paradigm for overcoming the limited context windows of large models, such as Large Language Models (LLMs) and Large Vision-Language Models (LVLMs), by incorporating external retrieval~\cite{zhao2023retrieving,abootorabi2025ask,mei2025survey}. In multimodal document scenarios, existing RAG approaches typically follow five strategies (as shown in Figure~\ref{fig:task}): (a) $RAG^{txt}_{Dense}$: converting multimodal content into plain text via Optical Character Recognition (OCR)~\cite{smith2007overview} and applying textual chunk-based RAG with LLMs~\cite{zhang2025ocr} and dense retrievers (e.g., BGE-m3~\cite{chen2024bge}); (b) $RAG^{txt+i2t}_{Dense}$: generating image descriptions (e.g., Image2Text (i2t)~\cite{hossain2019comprehensive}) and fusing them with OCR text using document content extraction toolkits (e.g., MinerU~\cite{wang2024mineru}) for textual chunk-based RAG; (c) $MRAG^{txt+img}_{CLIP}$: encoding text and images into a shared embedding space via multimodal encoders (e.g., CLIP~\cite{radford2021learning}, SigLIP~\cite{zhai2023sigmoid,tschannen2025siglip}) for retrieval followed by LVLM-based generation; (d) $VRAG^{page}_{ColPali}$: rendering document pages as images and retrieving relevant page-level content using vision-based methods (e.g., ColPali~\cite{faysse2024colpali}) prior to LVLM decoding; and (e) $GraphRAG^{txt+i2t}_{Dense}$: constructing knowledge graphs (KGs) from document content and applying Graph-based RAG~\cite{edge2024local,fan2025minirag}. However, these approaches often struggle to capture fine-grained cross-modal and cross-page associations, leading to incomplete grounding and suboptimal retrieval.

Document expansion methods such as Doc2Query~\citep{nogueira2019document} provide a principled way to map document content into a unified query representation space, enhancing retrieval by generating synthetic queries that capture a document’s latent information needs.
Building upon this idea, QCG-RAG~\citep{wu2025query} constructs a query-centric graph that explicitly links generated queries to their corresponding textual document chunks, enabling query-aware indexing and multi-hop retrieval over long textual contexts, thereby improving evidence aggregation and grounding in long-document question answering.
However, these approaches remain largely unexplored in the setting of multimodal long-context documents, where information is distributed across heterogeneous modalities and pages.

Building on this idea, we extend query-centric formulation to the multimodal setting and propose MLDocRAG (Multimodal Long-Context Document Retrieval-Augmented Generation), a framework for multimodal long-context document understanding. 
MLDocRAG leverages a unified retrieval structure—the Multimodal Chunk-Query Graph (MCQG)—constructed via MDoc2Query, which extends Doc2Query~\citep{nogueira2019document} to multimodal scenarios and generates semantically rich, answerable queries from heterogeneous chunks spanning text, images, and tables. 
The resulting MCQG links each query to its corresponding multimodal content, enabling selective retrieval and structured evidence aggregation across modalities and pages. 
This design effectively addresses cross-modal and cross-page associations, improving retrieval accuracy and grounding in multimodal long-document question answering (QA). 
Figure~\ref{fig:task}(f) illustrates the overall pipeline of our proposed MLDocRAG framework.
Our main contributions are as follows:

\noindent \textbullet\ We propose \textbf{MLDocRAG (Multimodal Long-
Context Document Retrieval-Augmented Generation)}, a unified framework for multimodal long-document QA that integrates multimodal document expansion with query-centric, graph-based retrieval for fine-grained and interpretable evidence selection. 

\noindent \textbullet\ We introduce the \textbf{MCQG (Multimodal Chunk-Query Graph)}, which links generated queries to corresponding multimodal chunks and connects semantically related information across pages.

\noindent \textbullet\ To construct MCQG, we leverage \textbf{MDoc2Query}, a multimodal document expansion process that generates answerable queries from multimodal chunks. 

\noindent \textbullet\ Extensive experiments on MMLongBench-Doc~\cite{ma2024mmlongbench} and LongDocURL~\cite{deng2025longdocurl} show that MLDocRAG consistently improves QA accuracy, advancing multimodal long-context document understanding.

\vspace{-0.5\baselineskip}
\section{Related Work}~\label{sect:related work}
\vspace{-1.5\baselineskip}

\pageenlarge{1} \looseness -1
\paragraph{Long-Context Document Understanding}
Understanding multimodal long-context documents, such as research papers or technical reports, requires resolving both cross-modal heterogeneity and long-range cross-page reasoning—capabilities still limited in current models~\cite{ ma2024mmlongbench,deng2025longdocurl}.  Despite the strong local alignment abilities of Large Vision-Language Models (LVLMs) like GPT-4o~\cite{hurst2024gpt}, Qwen2.5-VL~\cite{bai2025qwen2}, and Gemini~\cite{team2023gemini}, their fixed context windows limit their effectiveness in capturing globally relevant evidence dispersed across pages and modalities. This limitation leads to degraded performance in complex reasoning tasks, where key information is sparsely located, as highlighted in benchmarks such as MMLongBench-Doc~\cite{ma2024mmlongbench} and LongDocURL~\cite{deng2025longdocurl}. Retrieval-Augmented Generation (RAG) offers partial relief by introducing external memory, yet struggles with retrieving semantically aligned multimodal content at scale. 
Retrieval-Augmented Generation (RAG) partly mitigates this by introducing external memory, but struggles to retrieve and integrate semantically aligned multimodal information at scale. 
These challenges call for new retrieval and representation strategies tailored to multimodal long-document understanding.

\vspace{-0.5\baselineskip}
\paragraph{Multimodal RAG}
Recent efforts in multimodal  Retrieval Augmented Generation (RAG) have explored various strategies to adapt long-document understanding to the multimodal setting. Common approaches include OCR-based text extraction, image captioning fused with text chunks, shared embedding retrieval via multimodal encoders (e.g., CLIP~\cite{radford2021learning}, SigLIP~\cite{zhai2023sigmoid,tschannen2025siglip}), and vision-based page retrieval using rendered document images (e.g., ColPali~\cite{faysse2024colpali}). 
Some integrate structured knowledge via document-derived graphs to enhance reasoning~\cite{zhang2025comprehensive,zhao2023retrieving,abootorabi2025ask,mei2025survey}. However, existing pipelines operate at coarse granularity, overlooking fine-grained cross-modal and cross-page associations, leading to incomplete grounding and retrieval mismatches. This motivates developing semantically aligned and structurally informed retrieval for multimodal contexts.

\pageenlarge{1} \looseness -1
\vspace{-0.5\baselineskip}
\paragraph{Document Expansion}
Document expansion has been widely adopted in RAG settings to improve retrieval coverage by generating synthetic contents that anticipate potential information needs~\cite{tao2006language,singhal1999document,efron2012improving,wan2007single}. Methods such as Doc2Query~\citep{nogueira2019document} and its enhanced variant Doc2Query{-}{-}~\cite{gospodinov2023doc2query} generate diverse, semantically meaningful queries from document content, effectively enriching the retrieval index. Building on this idea, QCG-RAG~\citep{wu2025query} introduces a query-centric graph structure that connects queries to their source textual chunks, enabling multi-hop retrieval and structured evidence aggregation in long-document scenarios. 
Yet, these methods remain unimodal, lacking explicit modeling of cross-modal relationships. Extending query-centric expansion to multimodal documents remains an open problem for capturing fine-grained, heterogeneous associations.

\section{Methodology}~\label{sect:info source}
\vspace{-1\baselineskip}

\begin{figure*}[htbp]
  \centering
  \includegraphics[width=1.0\textwidth]{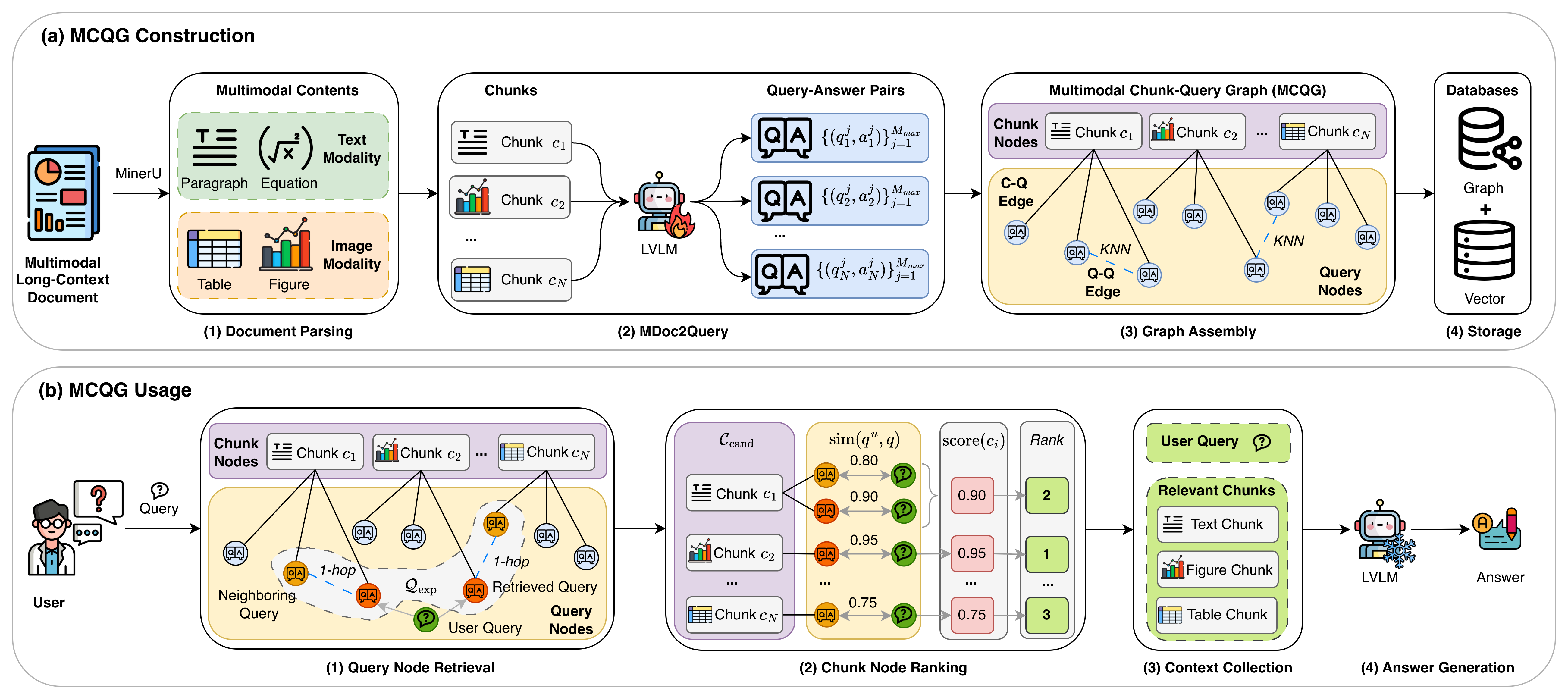}
  \caption{Overview of our proposed MLDocRAG framework, consisting of (a) \textit{MCQG Construction} for building a multimodal chuck-query graph and (b) \textit{MCQG Usage} for retrieving relevant multimodal chunks in long-document QA.}
  \label{fig:mldocrag}
  \vspace{-1\baselineskip}
\end{figure*}

\subsection{Preliminaries}

We consider the task of multimodal long-context document question answering, where the goal is to answer a user natural language question $q^u$ based on a multimodal long-context document $D$ that spans multiple pages and contains heterogeneous content. Each document $D$ originates from a PDF file and is processed into a sequence of pages $D=\{P_1, P_2, \ldots, P_X\}$ via OCR. 
Each page $P_x$ consists of a set of modality-specific chunks $C_i = \{c_1, c_2, \ldots, c_N\}$, where each chunk $c$ is associated with a modality (text ($txt$) or image ($img$)) and a content type (e.g., paragraph, figure, table, or equation).

Text chunks are contiguous paragraphs or extracted titles; image chunks consist of a visual region paired with an associated caption; table chunks include a caption, a rendered table image, and an OCR-converted Markdown-style textual representation~\cite{wang2024mineru}. The QA task follows a retrieval-augmented generation (RAG) paradigm: given a query $q$, a retriever selects the top-$K$ relevant multimodal chunks $\{c'_1, \ldots, c'_K\}$ from $D$, and a LVLM conditions on these chunks to generate the final answer $\hat{a}$. The primary challenge lies in retrieving semantically aligned and cross-modally grounded chunks from the long, heterogeneous document to support accurate and coherent generation.

\subsection{Framework Overview}

To tackle the challenges of cross-modal heterogeneity and long-range reasoning in multimodal long-document QA, 
we propose a \textit{Multimodal Long-Context Document Retrieval Augmented Generation (MLDocRAG)} framework based on the construction and usage of a \textit{Multimodal Chunk-Query Graph (MCQG)} that organizes multimodal document content around semantically rich, answerable queries.
To construct MCQG, we move beyond conventional chunk-based retrieval paradigms that treat multimodal content as flat and independent units, and instead hypothesize that organizing multimodal long-context document understanding around generated, answerable queries enables more fine-grained, semantically aligned, and interpretable retrieval.
Inspired by prior work on document expansion via query generation (e.g., Doc2Query~\citep{nogueira2019document}), we extend this idea to the multimodal setting by proposing MDoc2Query—a multimodal document expansion framework that generates semantically rich queries from heterogeneous document chunks. 
These generated queries act as \textit{retrieval anchors} that bridge the gap between user information needs and multimodal document content.
Figure~\ref{fig:mldocrag} illustrates the overall architecture of our proposed MLDocRAG framework which consists of two main stages: \textit{MCQG Construction} and \textit{MCQG Usage}.

\paragraph{MCQG Construction.} 
In this stage, a multimodal long document parsed from a PDF is decomposed into modality-specific chunks, where text-modality chunks correspond to paragraph text (including equations), and image-modality chunks correspond to figures or tables, each associated with its caption and any OCR-derived structured textual content.
We then apply the MDoc2Query process to generate a set of answerable queries for each chunk using a Large Vision-Language Model (LVLM). 
These queries are explicitly linked to their source chunks and further connected to semantically similar queries across the document, forming the \textit{Multimodal Chunk-Query Graph (MCQG)}. This graph captures both intra-modal and inter-modal associations, 
and provides a structured retrieval index that aligns semantically meaningful questions with relevant multimodal evidence.

\paragraph{MCQG Usage.} 
At inference time, a user query is embedded and matched against nodes in the MCQG using KNN-based retrieval in the query embedding space. 
By retrieving semantically similar generated queries and aggregating their linked source chunks, 
the system collects a compact yet relevant set of multimodal content. 
The retrieved chunks are further ranked based on their semantic similarity to the user query and 
provided as context to a Large Language or Vision-Language Model (LLM/LVLM) for final answer generation.  
This query-centric retrieval strategy enables interpretable, structured evidence aggregation over multimodal long contexts.

\subsection{MCQG Construction}

The MCQG Construction stage aims to transform a multimodal long-context document into a query-centric retrieval structure that supports fine-grained, semantically aligned, and cross-modal evidence retrieval. Specifically, we convert the long document into a collection of multimodal chunks, generate answerable queries from each chunk, and construct a graph that encodes chunk-query and inter-query associations. This process comprises four steps: (1) Document Parsing, (2) MDoc2Query, (3) Graph Assembly, and (4) Vector \& Graph Storage.

\paragraph{(1) Document Parsing.}  
We adopt an existing multimodal PDF parsing tool, such as \texttt{MinerU}~\cite{wang2024mineru}, to extract structured information from the document. Given a PDF document $D = \{P_1, P_2, \dots, P_X\}$ of $X$ pages, we extract a layout-preserving sequence of heterogeneous document components, including:
(i) paragraphs (text only);
(ii) figures (images with captions);
(iii) tables (table images accompanied by textual captions and OCR-converted Markdown text); and
(iv) equations (parsed into Markdown text).
The extracted content is stored in a JSON format ordered by visual layout position.
Based on this structured output, we define a set of modality-specific chunks:
\begin{equation}
    \mathcal{C} = \{c_1, c_2, \dots, c_N\}, \quad c_i \in \{\texttt{text}, \texttt{image}\}.
\end{equation}
Paragraph text is segmented into overlapping spans using a sliding window with a maximum token length and a fixed stride.
Equations are treated as part of the regular text content.
In contrast, each figure or table is treated as an image-modality chunk, while preserving its associated caption and any structured OCR-derived textual content.
Visual noise filtering can be further applied to remove uninformative images (e.g., blank pages, string-only images, or logos; see Appendix~\ref{appendix:visual_filtering}) via visual chunk classification using zero-shot CLIP inference.

\paragraph{(2) MDoc2Query.}  
To bridge document content with potential information needs, we extend the Doc2Query paradigm to multimodal settings. For each chunk $c_i \in \mathcal{C}$, we employ a Large Vision-Language Model (LVLM) to generate a set of answerable query–answer pairs:
\begin{equation}
    \mathcal{Q}_i = \left\{(q_{i}^{(1)}, a_{i}^{(1)}), \dots, (q_{i}^{(M_i)}, a_{i}^{(M_i)})\right\},
\quad M_i \leq M_{\text{max}}.
\end{equation}
Here, $q_{i}^{(j)}$ denotes a generated query and $a_{i}^{(j)}$ its corresponding answer, both grounded in chunk $c_i$. The number of pairs per chunk is adaptively determined by the richness of its content, up to a global cap $M_{\text{max}}$ per chunk.
Each query–answer pair $(q_{i}^{(j)},a_{i}^{(j)})$ is embedded into a dense vector representation using a pretrained text encoder $\phi(\cdot)$ (e.g., BGE-m3~\cite{chen2024bge}):
\begin{equation}
    \mathbf{v}_{i}^{(j)} = \phi\left([q_{i}^{(j)} ; a_{i}^{(j)}]\right) \in \mathbb{R}^d,
\end{equation}
where $[q; a]$ denotes concatenation of the query and answer as the retrieval unit.  
We simplify notation in the remainder of the paper by referring to $q_{i}^{(j)}$ as a shorthand for the full query–answer pair. 

\paragraph{(3) Graph Assembly.}  
We build the Multimodal Chunk-Query Graph (MCQG) as a heterogeneous graph $\mathcal{G} = (\mathcal{V}, \mathcal{E})$ with nodes:
\begin{equation}
    \mathcal{V} = \mathcal{C} \cup \mathcal{Q},
\quad \text{where } \mathcal{Q} = \bigcup_{i=1}^N \mathcal{Q}_i.
\end{equation}
Nodes in $\mathcal{V}$ include \textbf{Chunk Nodes} $\mathcal{C}$ and \textbf{Query Nodes} $\mathcal{Q}$. Edges in $\mathcal{E}$ include:
(1) \textbf{Chunk–Query (C-Q) Edges}: Each query $q_{i}^{(j)}$ is connected to its originating chunk $c_i$ via a directed anchor edge.
(2) \textbf{Query–Query (Q-Q) Edges}: We compute semantic similarity between queries using inner product of embeddings and connect each query to its top-$k$ nearest neighbors (KNN):
\begin{equation}
    \text{sim}(q, q') = \langle \phi([q;a]), \phi([q';a']) \rangle + \epsilon.
\end{equation}
Here, $q$ and $q'$ denote two generated queries. $\langle \cdot, \cdot \rangle$ denotes the inner product.
All query embeddings are $\ell_2$-normalized prior to similarity computation.
The constant offset $\epsilon$ (e.g., $\epsilon=1.0$) is added to ensure non-negative similarity scores for stable KNN construction.
This dual-edge structure allows the graph to capture both local chunk associations and global semantic neighborhoods across queries, enabling multi-hop traversal and cross-modal reasoning during retrieval.

\paragraph{(4) Storage.}  
To support scalable and efficient retrieval, we decouple vector retrieval from graph traversal by storing:
(1) Query–Answer vectors $\{\mathbf{v}_{i}^{(j)}\}$ in a dense \textbf{vector database} (e.g., FAISS~\cite{douze2025faiss}, ElasticSearch ~\cite{kuc2013elasticsearch}), supporting fast approximate nearest neighbor (ANN) search~\cite{aumuller2020ann}.
(2) The full MCQG graph structure in a \textbf{graph database} (e.g., Neo4j~\cite{guia2017graph}), preserving chunk–query and query–query relationships.

Importantly, we do \emph{not} perform vectorization of multimodal chunks (e.g., figure or table content) directly. Instead, all retrieval operates in the query–answer space, thereby avoiding the need for complex multimodal embedding alignment and reducing storage and computation overhead. The generated queries serve as interpretable, semantically rich anchors that effectively summarize and index the multimodal document content.

\subsection{MCQG Usage}

The MCQG Usage stage aims to retrieve semantically relevant multimodal content in response to a user query by leveraging the structure of the Multimodal Chunk-Query Graph (MCQG). Rather than retrieving from raw document chunks, our MLDocRAG approach retrieves and aggregates content via generated queries that serve as semantically aligned retrieval anchors. This process involves four main steps: (1) Query Node Retrieval, (2) Chunk Node Ranking, (3) Context Collection, and (4) Answer Generation.

\paragraph{(1) Query Node Retrieval.}  
Given a user query $q^u$, we first compute its vector embedding $\phi(q^u)$ using the same query encoder used during MCQG construction. 
We perform approximate nearest neighbor (ANN) search over the query–answer vectors in the vector database to retrieve the top-$n$ semantically similar generated queries:
\begin{equation}
    \mathcal{Q}_{\text{ret}} =
    \mathrm{Top}_n
    \left\{
    q \in \mathcal{Q}
    \;\middle|\;
    \mathrm{sim}(q^u, q) \ge \alpha
    \right\}.
\end{equation}
In practice, the number of retrieved queries is jointly constrained by the maximum node budget $n$ and a similarity threshold $\alpha \in [0, 2]$ (e.g., $\alpha = 1.0$), such that only queries with $\mathrm{sim}(q^u, q) \ge \alpha$ are retained.
To capture broader contextual evidence, we expand each retrieved query node $q \in \mathcal{Q}_{\text{ret}}$ in the MCQG via multi-hop neighbor expansion. For each query $q \in \mathcal{Q}_{\text{ret}}$, we traverse $h$ hops in the graph (through Query–Query Edges) and collect its neighboring queries:
\begin{equation}
    \mathcal{Q}_{\text{exp}} = \mathcal{Q}_{\text{ret}} \cup \bigcup_{q \in \mathcal{Q}_{\text{ret}}} \text{Nbr}_h(q),
\end{equation}
where $\text{Nbr}_h(q)$ denotes the set of $h$-hop neighbors of $q$ in the MCQG, and $\mathcal{Q}_{\text{exp}}$ represents the expanded query set that augments the initially retrieved queries with semantically related queries discovered through multi-hop graph traversal.

\paragraph{(2) Chunk Node Ranking.}
Each query $q \in \mathcal{Q}_{\text{exp}}$ is linked to a source multimodal chunk $c \in \mathcal{C}$. 
We collect all chunks associated with the expanded query set:
\begin{equation}
    \mathcal{C}_{\text{cand}} = \left\{ c_i \mid \exists\, q \in \mathcal{Q}_{\text{exp}} \ \text{s.t.}\ (q, c_i) \in \mathcal{E} \right\}.
\end{equation}
To prioritize the most relevant evidence, we assign a relevance score to each candidate chunk $c_i \in \mathcal{C}_{\text{cand}}$ based on the maximum semantic similarity between the user query $q^u$ and any query linked to $c_i$:
\begin{equation}
    \mathrm{score}(c_i) = 
    \max_{\substack{q \in \mathcal{Q}_{\text{exp}} \\ (q,c_i)\in \mathcal{E}}}
    \; \mathrm{sim}\bigl(q^u, q\bigr).
\end{equation}
Here, $\{ q \mid (q,c_i)\in\mathcal{E} \}$ denotes the subset of expanded queries in $\mathcal{Q}_{\text{exp}}$ that are connected to chunk $c_i$ in the MCQG.

\paragraph{(3) Context Collection.}
Based on the relevance scores, we select the top-$K$ (e.g., $K=5$) ranked multimodal chunks:
\begin{equation}
    \mathcal{C}_{\text{rel}} = \mathrm{Top}_K \left\{ c_i \in \mathcal{C}_{\text{cand}} \right\},
\end{equation}
where the ranking is determined by $\mathrm{score}(c_i)$.
The selected chunks are concatenated to form the multimodal retrieval context for the LVLM.
These chunks may include text blocks, image regions with captions, and tables augmented with structured and OCR-derived textual content.

\paragraph{(4) Answer Generation.}
Finally, the selected multimodal context $\mathcal{C}_{\text{rel}}$ is provided to a Large Vision--Language Model (LVLM), together with the original user query $q^u$, to generate the final answer:
\begin{equation}
    \hat{a} = \mathrm{LVLM}\!\left(q^u, \mathcal{C}_{\text{rel}}\right).
\end{equation}
This generation step benefits from the query-centric retrieval strategy and graph-based evidence aggregation, which together improve grounding, coverage, and factual consistency when answering complex questions over multimodal long-context documents.

\subsection{MDoc2Query Optimization}

The effectiveness of MLDocRAG largely depends on the quality of the Multimodal Chunk–Query Graph (MCQG), which is constructed via MDoc2Query.
In particular, the quality and granularity of the generated \emph{answerable queries} produced by MDoc2Query are critical, as they directly determine the semantic coverage, retrievability, and grounding fidelity of the overall pipeline.
To this end, we explore optimization strategies for MDoc2Query from both \emph{non-parametric} and \emph{parametric} perspectives.

\paragraph{Non-Parametric Optimization}
By default, MDoc2Query in MLDocRAG employs a LVLM to generate a set of answerable queries from parsed document chunks, such as cropped images or parsed table segments.
However, document parsing tools (e.g., MinerU~\cite{wang2024mineru}) often strip away essential contextual information—including figure captions, table headers, and hierarchical section titles—resulting in isolated chunks that can be semantically ambiguous and consequently degrade the quality of the generated queries.
To mitigate this issue, we adopt a \textit{Page-Context-Aware Generation} strategy.
Specifically, for a given chunk $c_i$ located on page $x$ with the corresponding page rendering image $P^{\text{page}}_x$, we construct the LVLM input as a tuple $(c_i, P^{\text{page}}_x)$.
Incorporating page-level visual context enables the model to resolve ambiguities arising from incomplete local information, such as coreference resolution (e.g., linking a chunk labeled ``Table~3'' to its corresponding description on the same page).

\paragraph{Parametric Optimization}
In addition to non-parametric strategies, we further explore parametric optimization by fine-tuning a pretrained Large Vision--Language Model (LVLM) on a curated set of high-quality multimodal \emph{chunk-to-query} exemplars.
Each training instance consists of a multimodal chunk $c$ (including text, an image with its caption, or a table with structured content) paired with a set of human-curated or automatically synthesized answerable query–answer $(q,a)$ pairs.
The LVLM is trained using standard teacher forcing, where the model conditions on the input chunk to generate the corresponding answer and subsequently autoregressively decodes the associated queries.
Through parametric adaptation, the model learns to produce more semantically precise, context-aware, and structurally grounded queries, thereby improving the expressiveness and reliability of MDoc2Query for downstream MCQG construction.

\section{Experimental Setup}~\label{sect:exp_setup}
In this section, we evaluate the effectiveness of the proposed MLDocRAG framework for multimodal long-context document question answering (QA), and compare it with existing approaches (as illustrated in Figure~\ref{fig:task}).
Specifically, our experimental study is designed to address the following research questions:

\noindent \textbullet\ \textbf{RQ1:} How does MLDocRAG perform on multimodal long-context document QA compared with baseline methods?

\noindent \textbullet\ \textbf{RQ2:} What are the effects on MLDocRAG of different MCQG node variants, including query node choices (query vs.\ answer), chunk ranking strategies (max vs.\ mean), and visual noise filtering?

\noindent \textbullet\ \textbf{RQ3:} How do key hyperparameters of MCQG usage affect the performance of MLDocRAG, such as expansion hops $h$, KNN neighbors $k$, and max nodes $n$?

\noindent \textbullet\ \textbf{RQ4:} What is the impact of MDoc2Query optimization from both non-parametric and parametric perspectives?

\subsection{Datasets \& Metrics}

\paragraph{Datasets.} 
We evaluate our MLDocRAG on two multimodal long-context document QA benchmarks: (1) \textbf{MMLongBench-Doc}~\cite{ma2024mmlongbench}: A curated benchmark for multimodal long-document understanding, consisting of documents in PDF format with diverse content including text, images, tables, and charts. Questions are paired with answerable evidence spread across pages and modalities. (2) \textbf{LongDocURL}~\cite{deng2025longdocurl}: A newly collected dataset containing web-based scientific and technical documents. Each instance includes a document (in PDF or HTML format), a natural language question, and annotated answerable segments across multimodal content.

\paragraph{Metrics.}
We adopt exact match Accuracy as the primary evaluation metric to measure the factual correctness of generated answers. To ensure consistent and scalable judgment across modalities and formats, we follow recent work and employ an \textit{LLM-as-a-Judge} protocol~\cite{gu2024survey}, using a strong LLM (e.g., Qwen2.5-72B~\cite{qwen2.5,qwen2}) to verify whether the predicted answer matches the gold reference answer. The evaluation prompt is shown in Appendix~\ref{appendix:llm_prompt}.

\subsection{Baselines}

We compare MLDocRAG with representative baselines from five categories:

\noindent \textbullet\  \textit{\textbf{Text Only (txt).}}
Use only text chunks with or without basic retrieval (BM25~\cite{robertson2009probabilistic} / BGE-m3~\cite{chen2024bge}):
$\text{LC}^{txt}$ with textual long-context reasoning, $\text{RAG}^{txt}_{\text{BM25}}$ with sparse retrieval, $\text{RAG}^{txt}_{\text{BGE-m3}}$ with dense retrieval.

\noindent \textbullet\  \textit{\textbf{Image2Text (txt+i2t).}}
Augment text with LVLM-generated image captions, treated as plain text: 
$\text{LC}^{txt+i2t}$, $\text{RAG}^{txt+i2t}_{\text{BM25}}$, $\text{RAG}^{txt+i2t}_{\text{BGE-m3}}$.

\noindent \textbullet\  \textit{\textbf{Multimodal (txt+img).}}
Encode both text and image chunks using multimodal embedding models for dense retrieval: 
$\text{LC}^{txt+img}$ for multimodal long-context reasoning, $\text{MRAG}^{txt+img}_{\text{CLIP}}$ with CLIP~\cite{radford2021learning}, $\text{MRAG}^{txt+img}_{\text{SigLIP}}$ with SigLIP~\cite{zhai2023sigmoid,tschannen2025siglip}, $\text{MRAG}^{txt+img}_{\text{ColPali}}$ with ColPali~\cite{faysse2024colpali}.

\noindent \textbullet\  \textit{\textbf{Page-level (page).}}
Render full document pages as images and perform vision-only retrieval followed by VQA: 
$\text{LC}^{page}$ for visual long-context reasoning, $\text{VRAG}^{page}_{\text{CLIP}}$, $\text{VRAG}^{page}_{\text{SigLIP}}$, $\text{VRAG}^{page}_{\text{ColPali}}$.

\noindent \textbullet\  \textit{\textbf{Graph.}} 
Construct knowledge graphs from extracted multimodal entities with the augmented text (i.e., \textit{txt+i2t}) for graph-based retrieval: $\text{GraphRAG}^{txt+i2t}_{BGE-m3}$ extended with efficient and lightweight MiniRAG~\cite{fan2025minirag}.

In addition, we evaluate several variants of the proposed MLDocRAG framework under different MCQG construction and usage settings to analyze the effects of key components and hyperparameters:
\textbf{(1) Node Variants.}  
(i) \textit{MLDocRAG w/ Query} to use queries only as nodes;
(ii) \textit{MLDocRAG w/ Answer} to use answers only as nodes;
(iii) \textit{MLDocRAG w/ Mean} to apply mean semantic similarity for chunk node ranking instead of the max operation;
(iv) \textit{MLDocRAG w/o Filter} to disable visual noise filtering during document parsing.
\textbf{(2) Hyperparameters.}  
(i) \textit{Hops} $h \in \{0, 1, 2, 3\}$ for multi-hop neighbor expansion;  
(ii) \textit{KNN} $k \in \{1, 2, 3, 4, 5\}$ for query-query edges;  
(iii) \textit{Max Nodes} $n \in \{5, 10, 15, 20\}$ for query node retrieval.
We additionally compare different query node retrieval backends (BGE-m3 vs.\ BM25) and similarity thresholds $\alpha \in \{1.0, 1.2\}$.
Note that MLDocRAG performs chunk-only query generation by default, whereas $\text{MLDocRAG}^{P}$ and $\text{MLDocRAG}^{P}_{P}$ additionally incorporate page context during query generation and final answer generation, respectively.

\subsection{Setup Details}~\label{subsect:setup_details}

\paragraph{Document Parsing Setting.}
Multimodal long-context documents are parsed using MinerU~\cite{wang2024mineru}, which extracts layout-ordered elements from PDF into structured JSON files. 
Chunks are constructed as follows:
(1) Text: Paragraphs are segmented using a maximum length of 1200 tokens with an overlap of 100 tokens. Equations parsed as Markdown are treated as regular text.
(2) Image: Figures and tables, together with their captions and OCR-derived content, are treated as individual image-modality chunks. In addition, document pages are rendered as images for page-level methods and MDoc2Query optimization.
We further apply visual noise filtering to image chunks via zero-shot classification using CLIP\footnote{\url{https://huggingface.co/openai/clip-vit-base-patch32}}, with details in Appendix~\ref{appendix:visual_filtering}.

\paragraph{Model Configuration.}
We employ the BGE-m3 encoder to generate dense embeddings for queries, which are stored and indexed in ElasticSearch ~\cite{kuc2013elasticsearch} as the core vector database to support efficient similarity search.
For query generation and final answer generation, we use \texttt{Qwen2.5-VL-32B}\footnote{\url{https://huggingface.co/Qwen/Qwen2.5-VL-32B-Instruct}}~\cite{bai2025qwen2.5-vl} by default, while \texttt{Qwen2.5-VL-7B}\footnote{\url{https://huggingface.co/Qwen/Qwen2.5-VL-7B-Instruct}}~\cite{bai2025qwen2.5-vl} is adopted for MDoc2Query optimization.
For evaluation, we adopt an LLM-as-a-Judge setup using \texttt{Qwen2.5-72B}\footnote{\url{https://huggingface.co/Qwen/Qwen2.5-72B-Instruct}}~\cite{qwen2.5,qwen2}.
All LLMs/LVLMs are deployed via the \texttt{SGLang}~\cite{zheng2024sglang} framework on NVIDIA H20 GPUs to enable high-throughput inference.

\paragraph{Default Hyperparameters.}
For experiments on both MMLongBench-Doc and LongDocURL, we adopt a unified set of default hyperparameters.
Specifically, we set the KNN neighborhood size to $k=3$ for query--query edge construction, retrieve up to $n=10$ query nodes with a similarity threshold $\alpha=1.2$, and perform $h=2$-hop query expansion.
For answer generation, the top-$K=5$ ranked multimodal chunks are selected as the retrieval context.

\section{Experimental Results}

In this section, we analyse the experimental results with respect to the four research questions stated in Section~\ref{sect:exp_setup} to gauge the effectiveness of our proposed MLDocRAG.

\begin{table*}[ht]
\footnotesize
\centering
\addtolength{\tabcolsep}{-3pt}
    \begin{tabular}{l|cc|cc|ccccc|ccc|c|cccc|ccc|c}
        \toprule
        & & & & & \multicolumn{9}{c|}{\textbf{MMLongBench-Doc}} & \multicolumn{8}{c}{\textbf{LongDocURL}} \\
        & \multicolumn{2}{c|}{Model} & \multicolumn{2}{c|}{Modality} & \multicolumn{5}{c|}{Evidence Source} & \multicolumn{3}{c|}{Evidence Page} & Overall & \multicolumn{4}{c|}{Evidence Source} & \multicolumn{3}{c|}{Evidence Page} & Overall \\
        Method & Retriever & GEN & TXT & IMG & TXT & LAY & CHA & TAB & FIG & SIN & MUL & UNA & ACC & TXT & LAY & FIG & TAB & SP & MP & CE & ACC \\
        \midrule
        LC$^{txt}$ & - & LLM & \ding{51} & \ding{55} & 44.3 & 31.9 & 20.8 & 12.8 & 22.0 & 32.8 & 21.9 & 69.3 & 36.9 & 59.1 & \underline{38.8} & 26.0 & 41.5 & 43.3 & 45.7 & \underline{29.2} & 40.2 \\
        RAG$^{txt}_{BM25}$ & BM25 & LLM & \ding{51} & \ding{55} & 44.9 & 26.1 & 20.2 & 14.2& 17.1 & 31.6 & 18.9 & 76.3 & 36.8 & 60.2 & 37.4 & 24.9 & 37.3 & 43.1 & 45.9 & 26.8 & 39.5  \\
        RAG$^{txt}_{BGE-m3}$ & BGE-m3 & LLM & \ding{51} & \ding{55} & 40.0 & 31.1 & 20.2 & 10.6 & 18.4 & 29.2 & 19.4 & \underline{77.2} & 36.0 & 59.8 & 34.7 & 19.3 & 31.5 & 41.4 & 44.6 & 21.0 & 36.8 \\
        \midrule
        LC$^{txt+i2t}$ & - & LLM & \ding{51} & \ding{55} & \underline{48.5} & 32.8 & 40.5 & 41.7 & 28.6 & 44.1 & \underline{32.5} & 54.8 & 42.5 & 51.1 & 31.0 & 29.1 & 33.6 & 42.2 & 49.0 & 15.7 & 37.3 \\
        RAG$^{txt+i2t}_{BM25}$ & BM25 & LLM & \ding{51} & \ding{55} & 43.6 & 31.9 & 38.2 & 38.1 & 27.0 & 45.3 & 26.9 & 66.7 & 43.7 & \underline{60.6} & 36.5 & 45.7 & 40.3 & \underline{61.9} & 54.1 & 22.4 & \underline{48.1} \\
        RAG$^{txt+i2t}_{BGE-m3}$ & BGE-m3 & LLM & \ding{51} & \ding{55} & 42.3 & 26.1 & 37.6 & 45.41 & 31.3 & 49.0 & 27.2 & 71.5 & \underline{46.5} & 59.7 & 32.8 & \underline{47.0} & \underline{42.3} & \underline{61.9} & \textbf{57.4} & 17.1 & 47.8 \\
        \midrule
        LC$^{txt+img}$ & - & LVLM  & \ding{51} & \ding{51} & \textbf{53.8} & \underline{34.5} & \textbf{45.5} & \underline{45.9} & \textbf{35.9} & \underline{52.4} & \textbf{37.8} & 45.6 & 46.1 & 57.7 & 34.7 & 30.9 & 39.8 & 46.0 & 48.9 & 23.4 & 40.7\\
        MRAG$^{txt+img}_{CLIP}$ & CLIP & LVLM  & \ding{51} & \ding{51} & 39.0 & 27.7 & 27.0 & 13.8 & 21.4 & 34.0 & 19.2 & 70.2 & 36.7 & 51.7 & 32.4 & 22.6 & 30.7 & 35.6 & 44.0 & 21.0 & 34.5 \\
        MRAG$^{txt+img}_{SigLIP}$ & SigLIP & LVLM  & \ding{51} & \ding{51} & 23.9 & 15.1 & 16.9 & 15.1 & 18.1 & 21.9 & 14.7 & \textbf{82.0} & 32.2 & 23.2 & 22.7 & 5.9 & 14.1 & 8.0 & 21.2 & 17.9 & 15.5 \\
        MRAG$^{txt+img}_{ColPali}$ & ColPali & LVLM  & \ding{51} & \ding{51} & 34.1 & 29.4 & 28.1 & 25.7 & 25.3 & 35.8 & 20.3 & 71.1 & 38.1 & 51.0 & 31.8 & 32.9 & 34.0 & 45.2 & 44.8 & 23.4 & 38.9 \\
        \midrule
        LC$^{page}$ & - & LVLM & \ding{55} & \ding{51} & 41.0 & 26.9 & 31.5 & 30.3 & 27.6 & 39.7 & 23.6 & 53.1 & 37.2 & 40.2 & 29.7 & 23.7 & 30.3 & 34.0 & 39.1 & 17.9 & 31.3  \\
        VRAG$^{page}_{CLIP}$ & CLIP & LVLM & \ding{55} & \ding{51} & 28.9 & 27.7 & 24.7 & 22.5 & 28.3 & 34.8 & 16.7 & 75.4 & 37.3 & 33.8 & 26.2 & 20.1 & 26.1 & 24.1 & 34.2 & 18.9 & 26.2 \\
        VRAG$^{page}_{SigLIP}$ & SigLIP & LVLM & \ding{55} & \ding{51} & 16.1 & 13.5 & 13.5 & 9.2 & 9.5 & 15.4 & 9.2 & \underline{77.2} & 26.3 & 21.2 & 25.3 & 17.0 & 19.1 & 17.4 & 17.7 & 25.8 & 19.9 \\
        VRAG$^{page}_{ColPali}$ & ColPali & LVLM & \ding{55} & \ding{51} & 40.0 & \textbf{37.8} & 32.0 & 30.7 & \underline{32.2} & 44.1 & 23.9 & 63.2 & 41.4 & 56.8 & \textbf{39.4} & 45.8 & \textbf{53.1} & 52.1 & 50.3 & \textbf{36.4} & 47.1  \\
        \midrule
        GraphRAG$^{txt+i2t}_{BGE-m3}$ & BGE-m3 & LLM & \ding{51} & \ding{55} &  48.1 & 25.5 & 36.4 & \textbf{51.5} & 30.2 & 51.3 & 27.1 & 48.5 & 42.6 & 58.2 & 33.9 & 32.8 & 36.1 & 59.2 & 54.4 & 15.7 & 45.3\\
        \midrule
        MLDocRAG & BGE-m3 & LVLM & \ding{51} & \ding{51} & 47.2 & \textbf{37.8} & \underline{42.7} & 41.3 & 31.9 & \textbf{52.6} & 26.4 & 71.5 & \textbf{47.9} & \textbf{65.1} & \textbf{39.4} & \textbf{48.3} & 41.1 & \textbf{66.9} & \underline{56.3} & 23.4 & \textbf{50.8} \\
        \bottomrule
    \end{tabular}
    \caption{Performance comparison on MMLongBench-Doc and LongDocURL measured by Accuracy (\%). For \textbf{MMLongBench-Doc}, five formats are considered: text (TXT), layout (LAY), chart (CHA), table (TAB), and image (FIG), with scopes including single-page (SIN), multi-page (MUL), and unanswerable (UNA). For \textbf{LongDocURL}, four formats are evaluated: text (TXT), layout (LAY), table (TAB), and figures (FIG), with evidence spans categorized as single-page (SP), multi-page (MP), or cross-element (CE), where CE denotes the integration of multiple modalities. Best results are shown in \textbf{bold}, and second-best results are \underline{underlined}.}
    \label{tab:results}
    \vspace{-2\baselineskip}
\end{table*}

\subsection{MLDocRAG vs. Baselines (RQ1)}

Table~\ref{tab:results} reports the performance of MLDocRAG and representative baselines on MMLongBench-Doc and LongDocURL in terms of accuracy (\%).
Overall, MLDocRAG achieves the best overall performance on both datasets, with an accuracy of 47.9\% on MMLongBench-Doc and 50.8\% on LongDocURL, consistently outperforming all baseline methods.
In particular, compared to text-only and image-to-text baselines, MLDocRAG benefits from explicitly modeling multimodal evidence without collapsing visual information into flat textual descriptions, thereby preserving fine-grained visual semantics that are critical for layout-, chart-, and figure-centric questions.
In contrast to multimodal dense retrieval methods that independently retrieve text and image chunks, MLDocRAG organizes multimodal content around generated, answerable queries and performs query-centric multi-hop expansion, enabling effective aggregation of semantically related evidence scattered across pages.
This advantage is particularly evident in multi-page settings, where simple chunk-level retrieval or page-level visual reasoning fails to capture long-range dependencies.
Moreover, in contrast to the text-only graph-based baseline, MLDocRAG explicitly models cross-modal associations through the Multimodal Chunk–Query Graph, allowing structured evidence propagation and ranking.
As a result, MLDocRAG delivers consistent gains across different evidence sources and document scopes, demonstrating superior grounding and robustness for multimodal long-context document question answering.

\begin{figure}[t]
\centering
\subfigure[MMLongBench-Doc]{
    \includegraphics[width=0.225\textwidth]{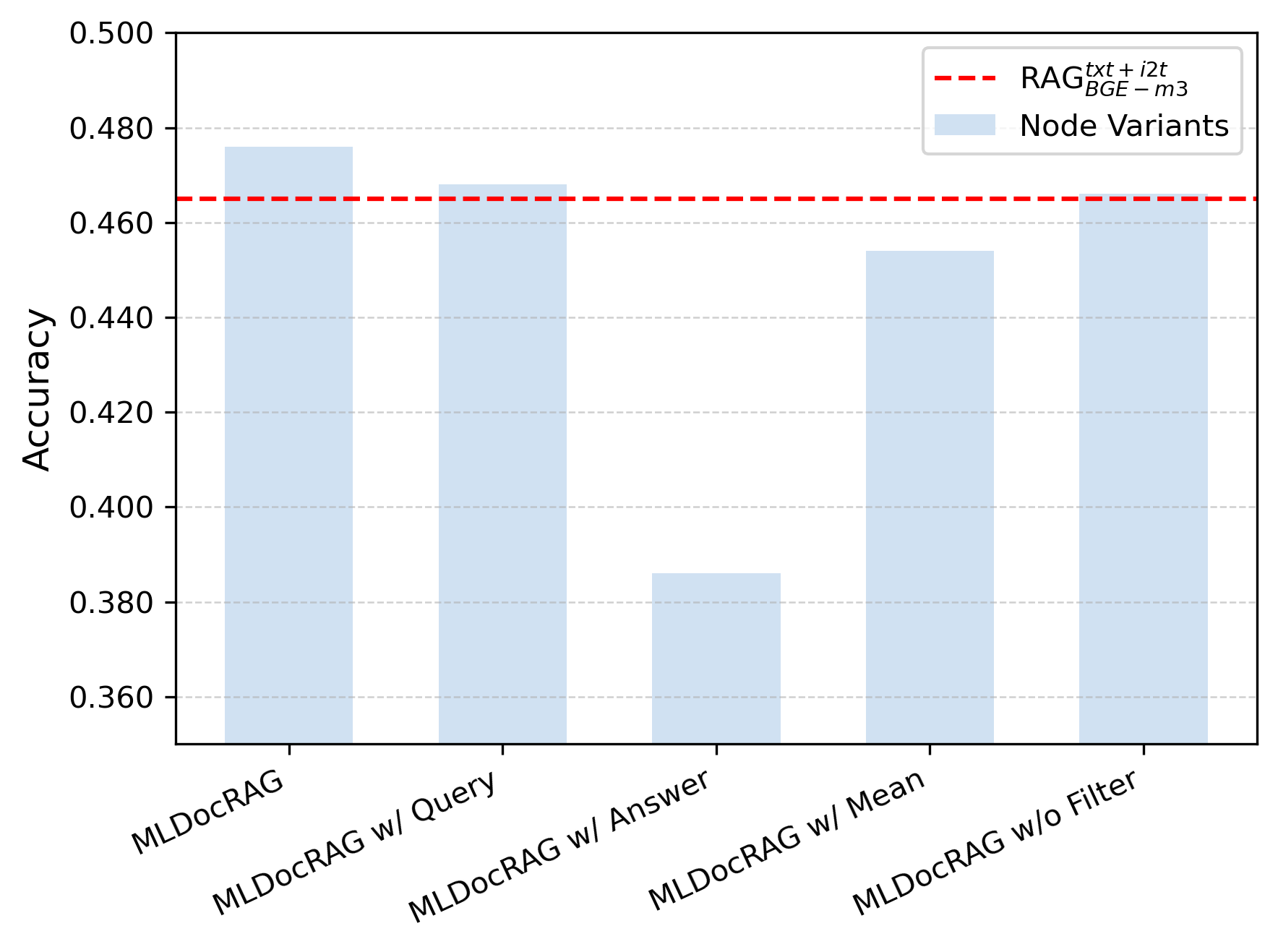}
}
\hfill
\subfigure[LongDocURL]{
    \includegraphics[width=0.225\textwidth]{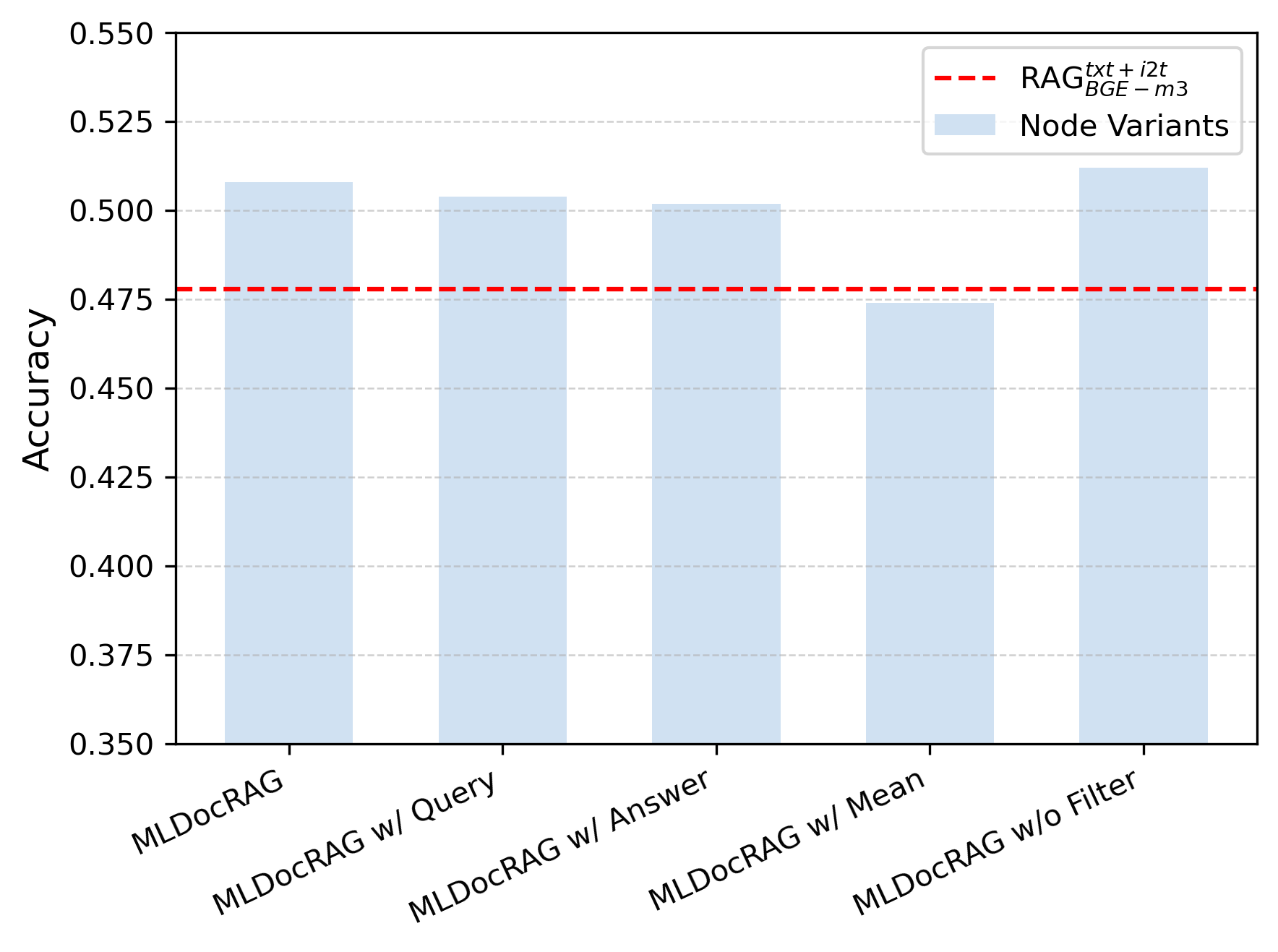}
}
\vspace{-1.0\baselineskip}
\caption{Ablation on node variants across both datasets.}
\label{fig:ablation_nodes}
\vspace{-1.0\baselineskip}
\end{figure}

\subsection{Impact of Node Variants (RQ2)}

Figure~\ref{fig:ablation_nodes} analyzes the impact of different MCQG node variants on the performance of MLDocRAG.
\textbf{(1) Query Node Choices (Query vs.\ Answer).}
Generated queries play a dominant role in performance.
While \textit{MLDocRAG w/ Query} remains relatively stable, \textit{MLDocRAG w/ Answer} exhibits substantial performance degradation (dropping below 39\% on MMLongBench-Doc), indicating that answers alone are insufficient as graph node representations.
The default MLDocRAG, which combines queries with answers, achieves the best results by leveraging answers as complementary contextual signals.
\textbf{(2) Chunk Node Ranking Strategies (Max vs.\ Mean).}
The Max strategy consistently outperforms Mean aggregation.
Using averaged similarity scores (\textit{MLDocRAG w/ Mean}) degrades performance compared to the default Max-based ranking, suggesting that a chunk’s relevance is better captured by its most relevant query rather than by an average over all associated queries.
\textbf{(3) Visual Noise Filtering.}
Visual noise filtering is crucial for effective retrieval.
Removing this component (\textit{MLDocRAG w/o Filter}) leads to noticeable performance drops, confirming that filtering out irrelevant image chunks (e.g., blank pages or logos) is necessary to prevent noise from interfering with retrieval.

\begin{figure}[tb]
\centering
\subfigure[Hops $h$]{
\begin{minipage}[t]{0.32\linewidth}
\centering
\includegraphics[width=1.0in]{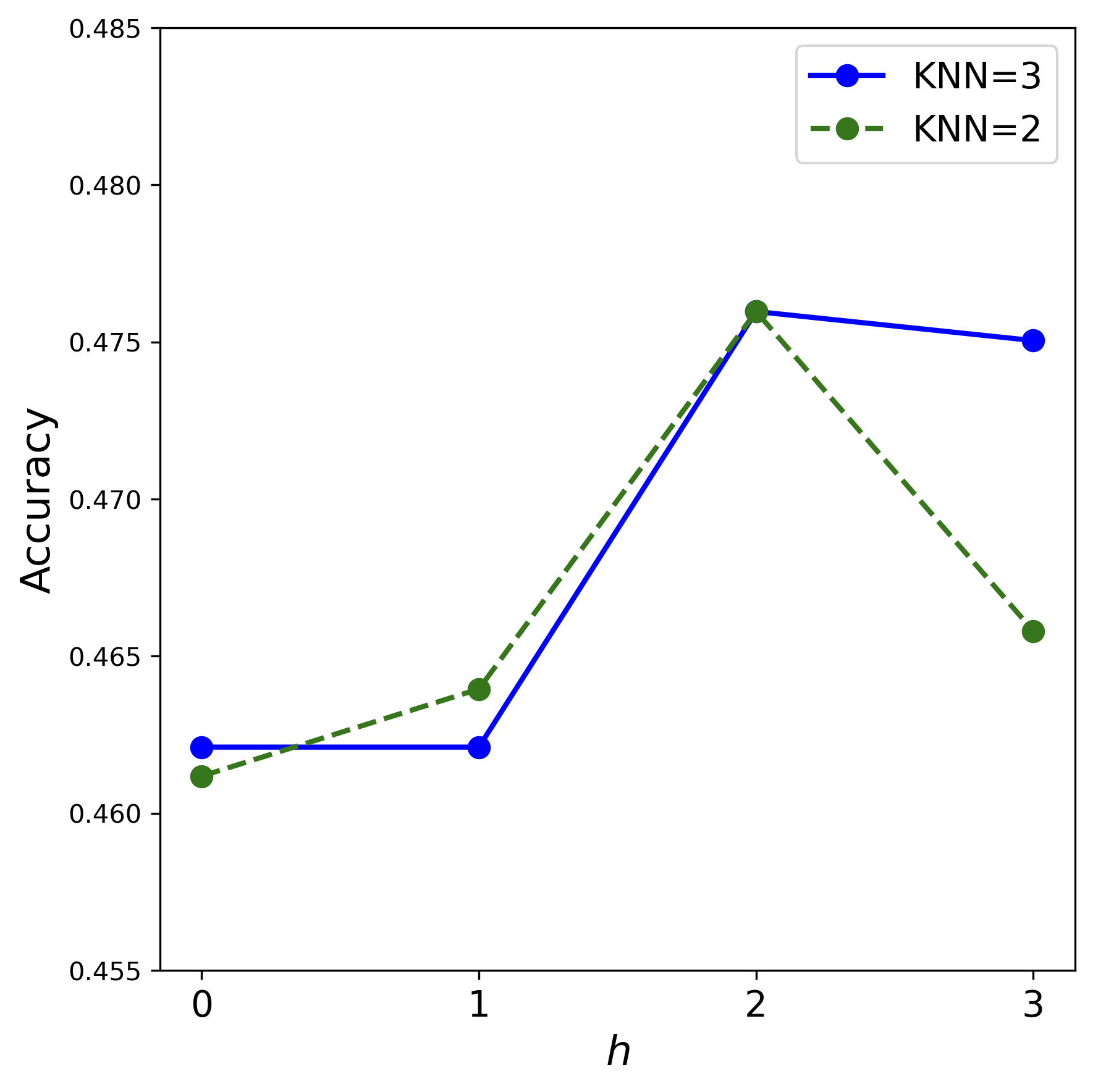}
\end{minipage}%
}%
\subfigure[KNN $k$]{
\begin{minipage}[t]{0.32\linewidth}
\centering
\includegraphics[width=1.0in]{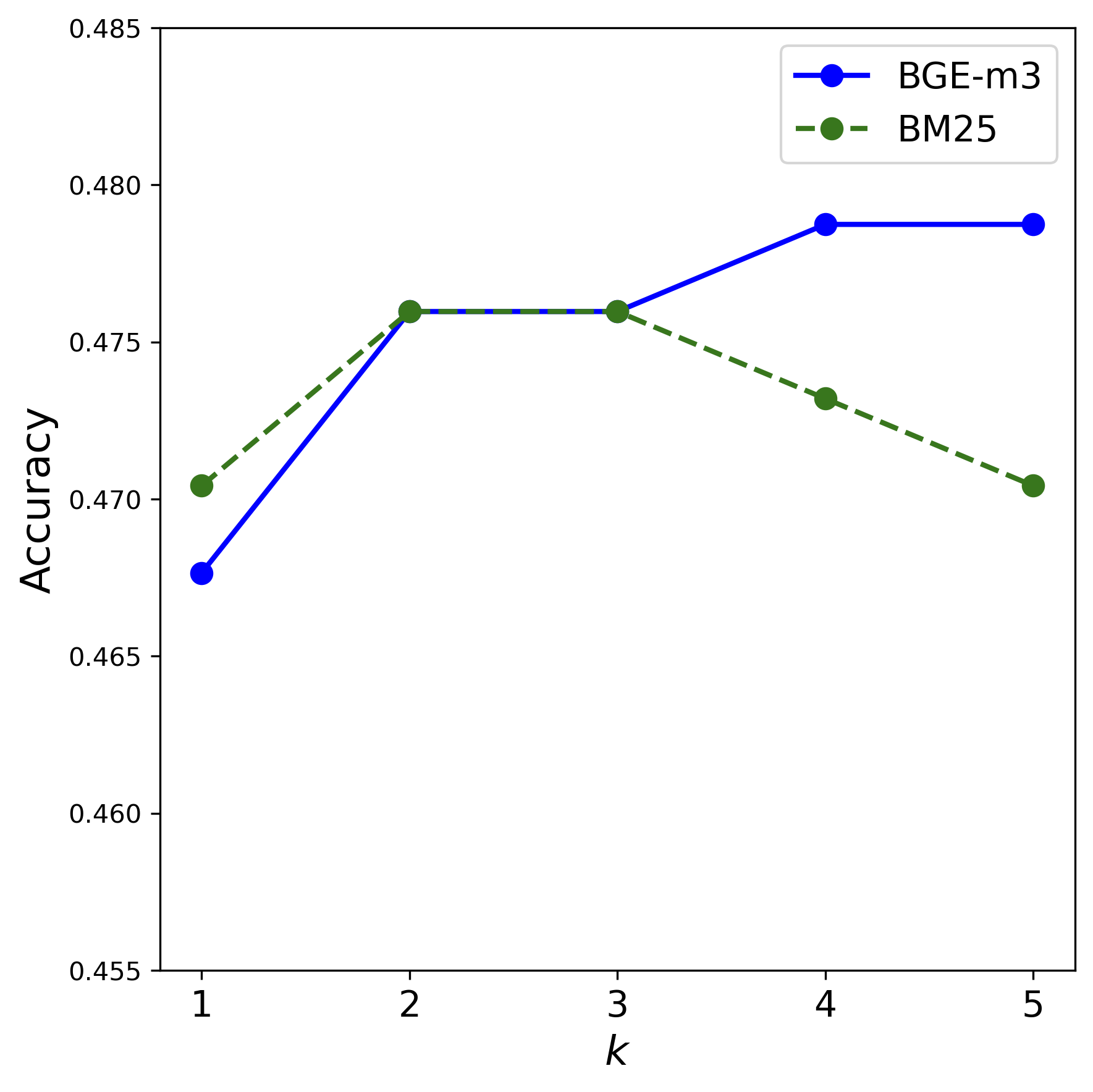}
\end{minipage}%
}%
\subfigure[Max Nodes $n$]{
\begin{minipage}[t]{0.32\linewidth}
\centering
\includegraphics[width=1.0in]{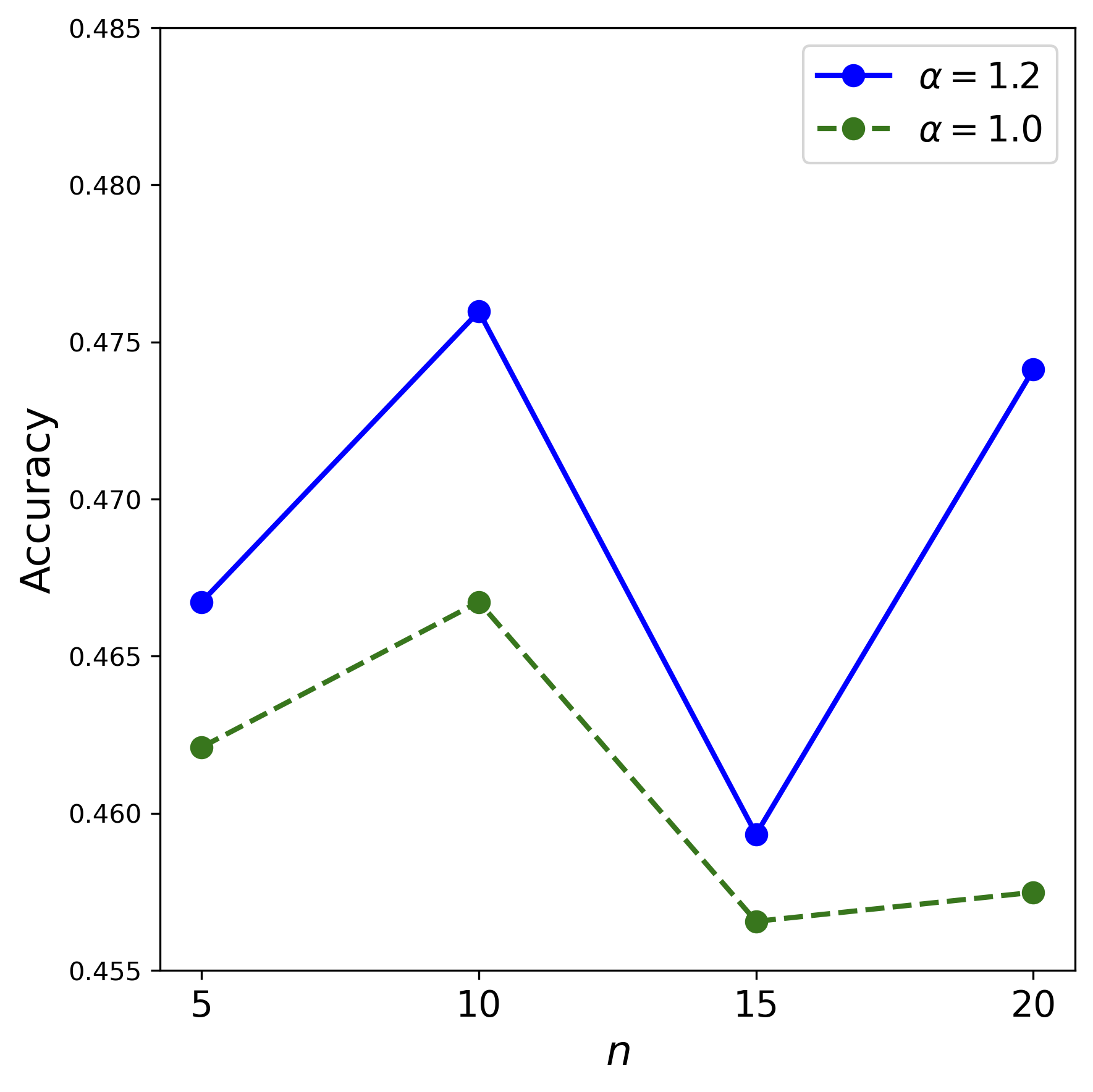}
\end{minipage}%
}%
\vspace{-1.\baselineskip}
\centering
\caption{Ablation on hyperparameters (MMLongBench-Doc).}
\label{fig:ablation-mmlongbenchdoc}
\vspace{-1\baselineskip}
\end{figure}

\begin{figure}[tb]
\centering
\subfigure[Hops $h$]{
\begin{minipage}[t]{0.32\linewidth}
\centering
\includegraphics[width=1.0in]{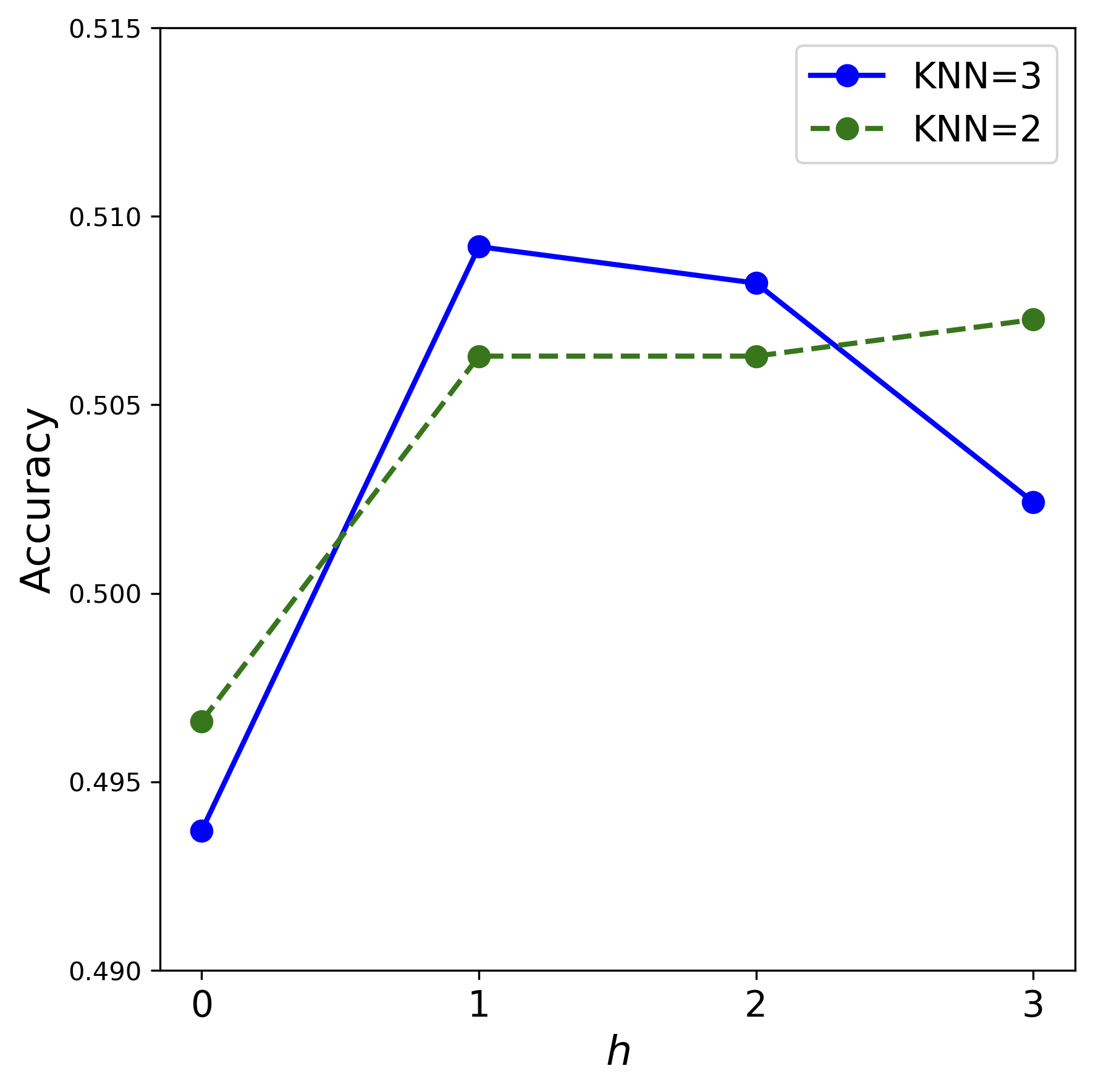}
\end{minipage}%
}%
\subfigure[KNN $k$]{
\begin{minipage}[t]{0.32\linewidth}
\centering
\includegraphics[width=1.0in]{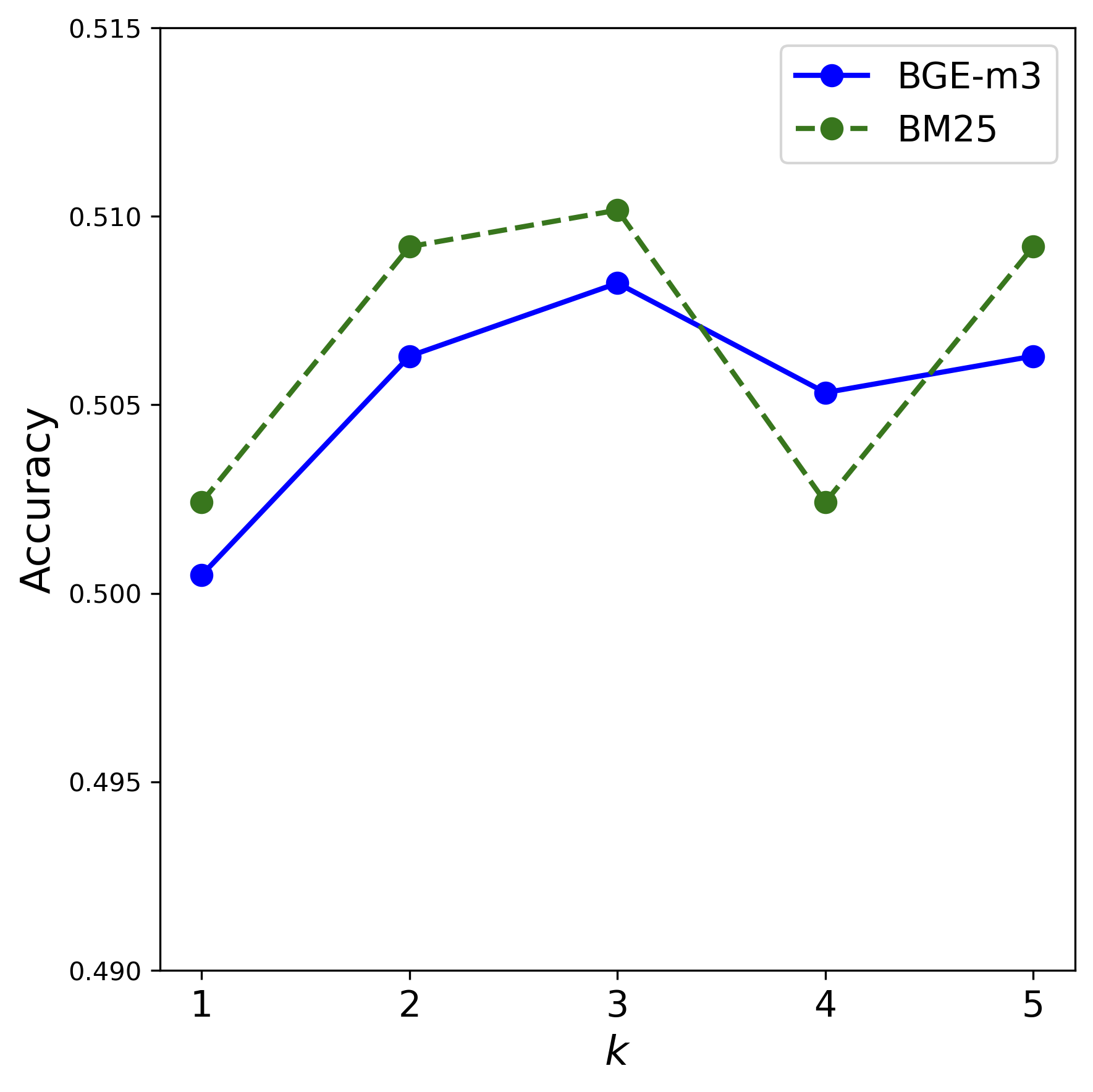}
\end{minipage}%
}%
\subfigure[Max Nodes $n$]{
\begin{minipage}[t]{0.32\linewidth}
\centering
\includegraphics[width=1.0in]{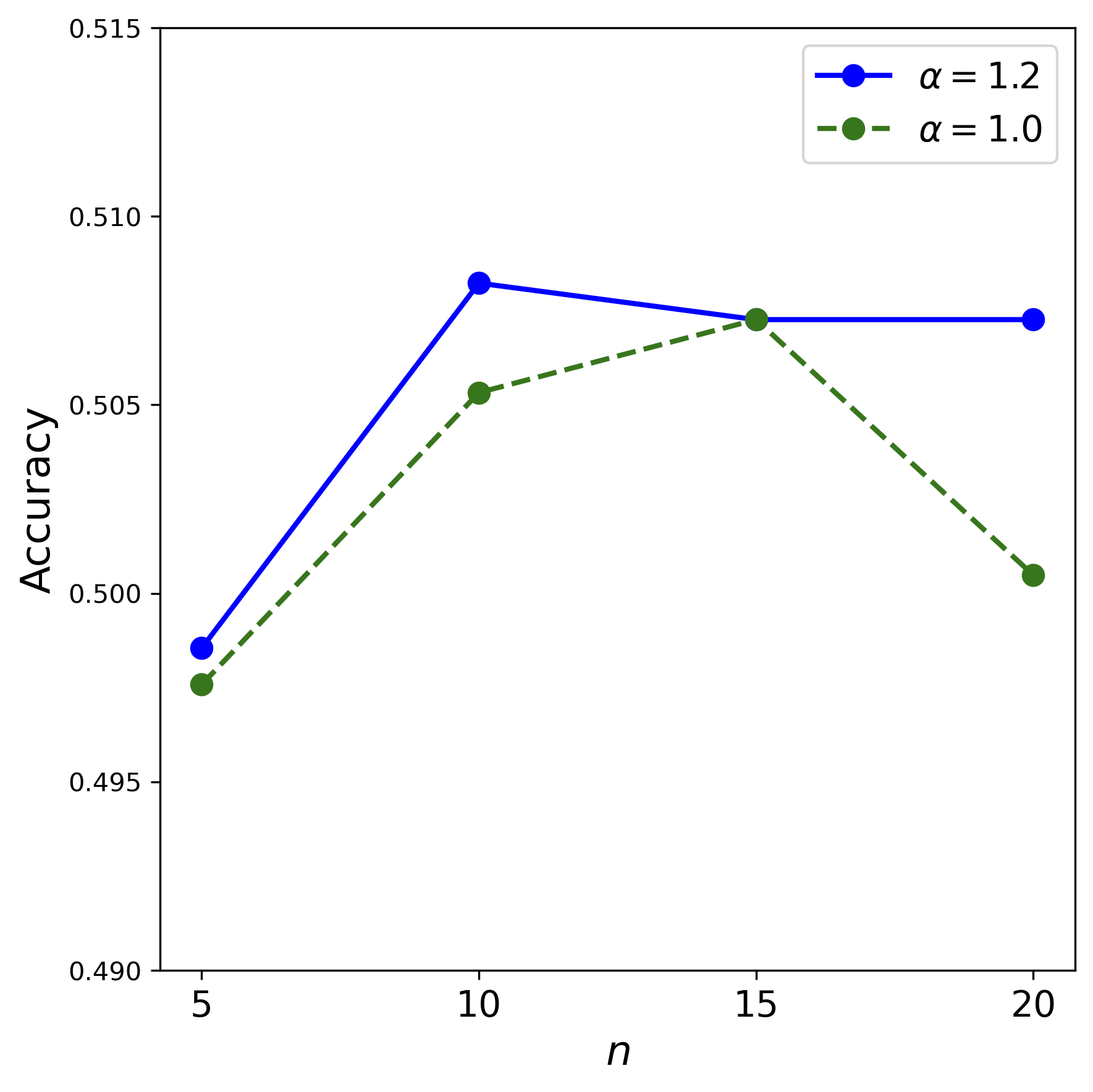}
\end{minipage}%
}%
\vspace{-1.\baselineskip}
\centering
\caption{Ablation on hyperparameters (LongDocURL).}
\label{fig:ablation-longdocurl}
\vspace{-1.\baselineskip}
\end{figure}

\subsection{Impact of Hyperparameters (RQ3)}

Figures~\ref{fig:ablation-mmlongbenchdoc} and~\ref{fig:ablation-longdocurl} illustrate the impact of key hyperparameters on the performance of MLDocRAG.
\textbf{(1) Expansion Hops ($h$).}
Multi-hop expansion ($h=1,2$) consistently outperforms zero-hop retrieval ($h=0$), highlighting the benefit of query node expansion through graph traversal.
However, performance degrades when $h=3$, indicating that excessive expansion introduces semantic noise that outweighs the gains from additional context.
\textbf{(2) KNN Neighbors ($k$).}
Accuracy generally improves as the number of nearest neighbors increases, peaking around $k=3$.
This suggests that a moderate graph density effectively bridges semantic gaps between queries, while denser connections beyond this point yield diminishing returns or introduce noise.
\textbf{(3) Max Nodes ($n$).}
Performance exhibits a clear optimum at $n=10$.
Retrieving too few nodes ($n=5$) limits evidence coverage, whereas retrieving too many nodes ($n>10$) introduces irrelevant information, leading to notable performance drops.
In addition, a stricter similarity threshold ($\alpha=1.2$) consistently outperforms a looser threshold ($\alpha=1.0$), emphasizing the importance of prioritizing high-quality entry nodes over quantity.

\begin{figure}[t]
\centering
\subfigure[MMLongBench-Doc]{
    \includegraphics[width=0.225\textwidth]{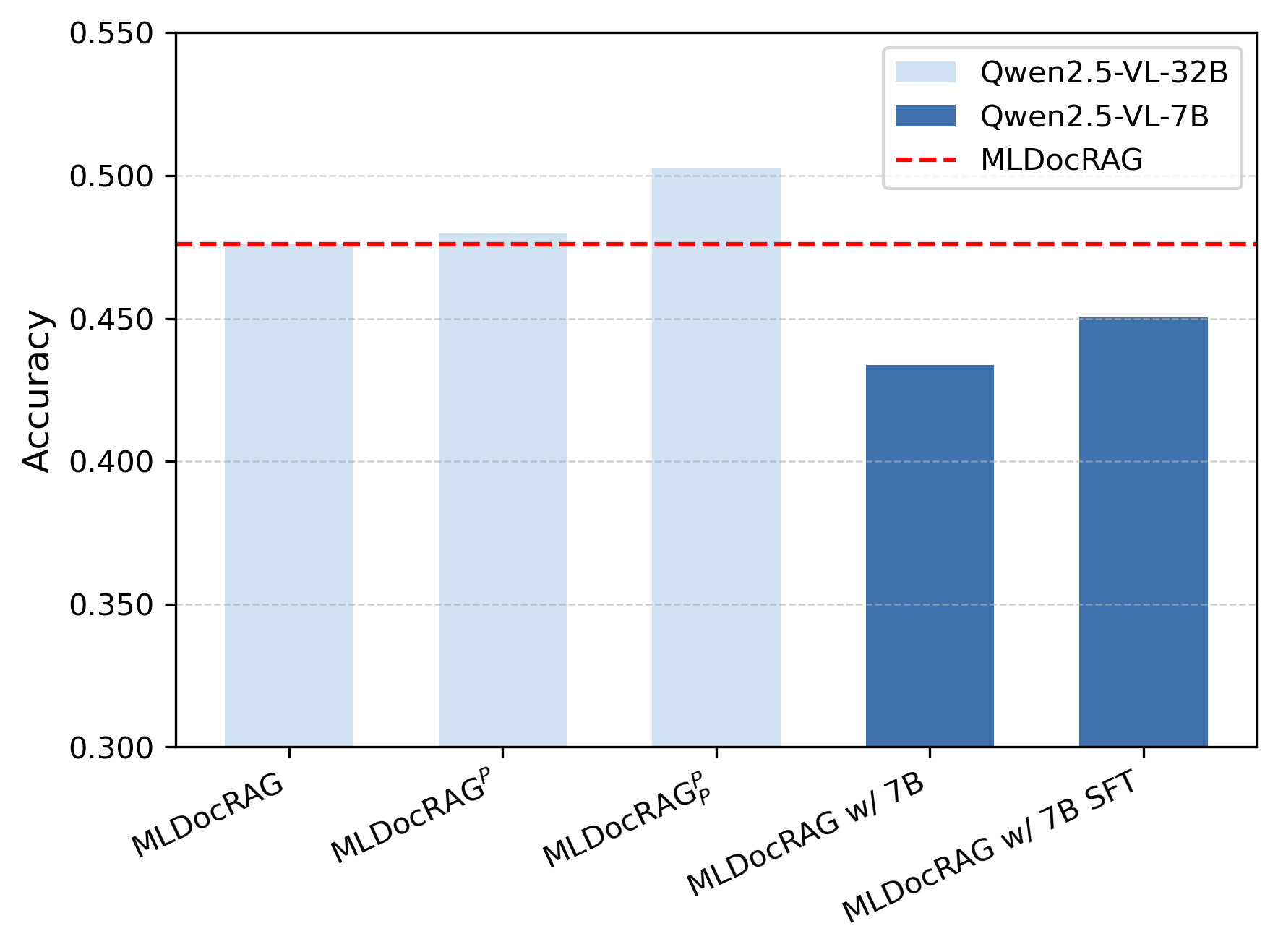}
}
\hfill
\subfigure[LongDocURL]{
    \includegraphics[width=0.225\textwidth]{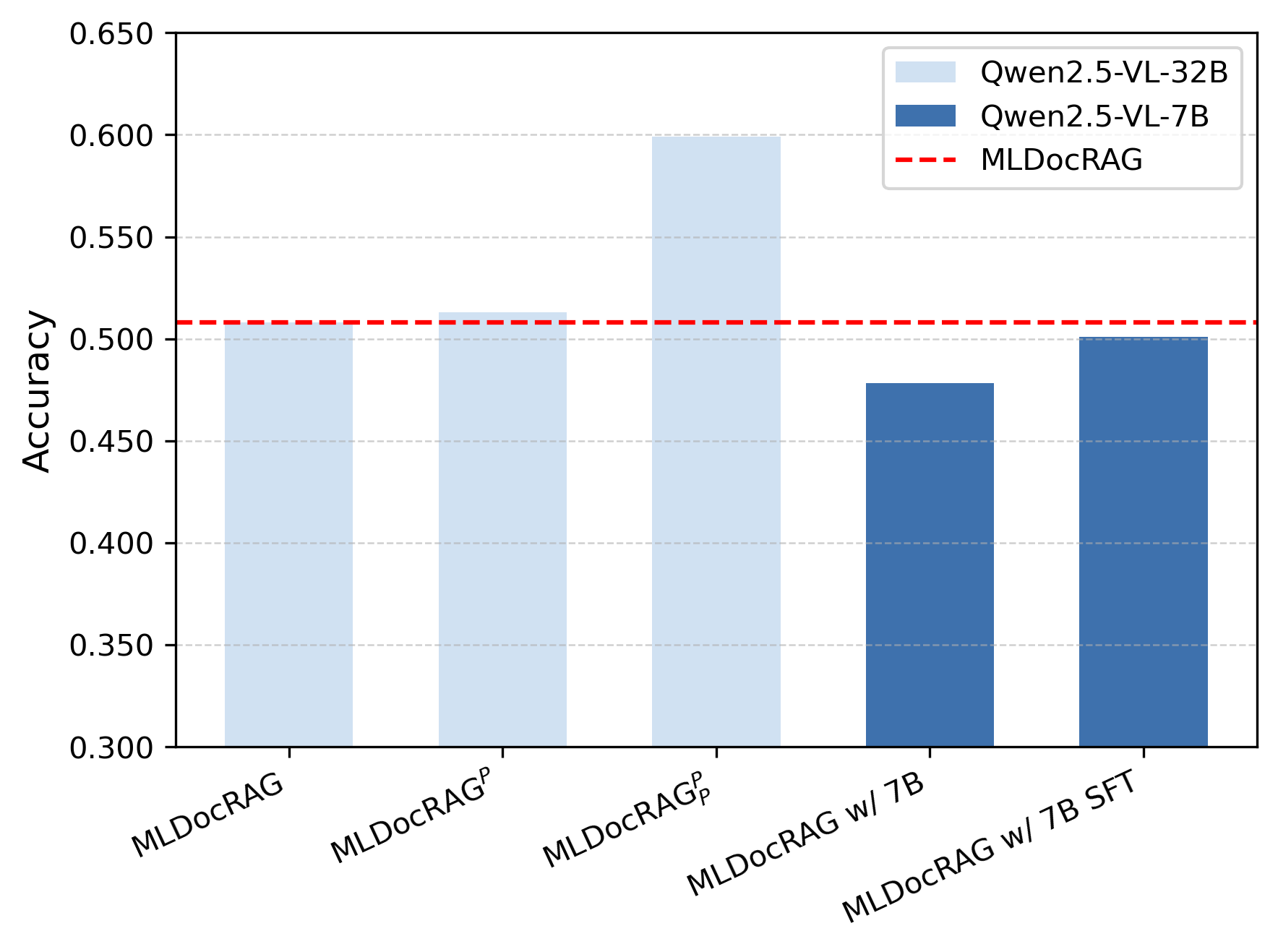}
}
\vspace{-1.0\baselineskip}
\caption{Ablation on MDoc2Query optimization.}
\label{fig:mdoc2query_optimization}
\vspace{-1.0\baselineskip}
\end{figure}

\subsection{Impact of MDoc2Query Optimization (RQ4)}

Figure~\ref{fig:mdoc2query_optimization} illustrates the impact of non-parametric and parametric optimization strategies for MDoc2Query on the performance of MLDocRAG.
\textbf{(1) Non-Parametric Optimization.}
Providing each chunk with its corresponding page image during query generation consistently improves performance.
Specifically, \textit{MLDocRAG$^{P}$} outperforms the chunk-only MLDocRAG across both datasets, demonstrating that incorporating page-level context enhances query quality by mitigating semantic fragmentation caused by isolated chunks.
Representative query--answer examples generated by MDoc2Query with and without page context are provided in Appendix~\ref{use case: mdoc2query example}.
Furthermore, \textit{MLDocRAG$^{P}_{P}$} integrates page-level information at both the query generation and answer generation stages, achieving the best overall performance.
This result confirms that supplying reference page context throughout both retrieval indexing and QA stages maximizes evidence utilization.
\textbf{(2) Parametric Optimization.}
MLDocRAG remains effective when deployed with smaller LVLMs.
Although the 32B model achieves the strongest overall performance, the 7B model with supervised fine-tuning (\textit{MLDocRAG w/ 7B SFT}) attains 50.1\% accuracy on LongDocURL, closely matching the 32B baseline (50.8\%).
These results demonstrate that MLDocRAG can be efficiently adapted to resource-constrained models through parametric optimization.

\begin{figure}[t!] 
\centering
\footnotesize
\begin{minipage}[t]{0.48\textwidth}
    \vspace{0pt} 
    \begin{modelbox}{white!10}{Question \& Answer \cmark}
    \textbf{Question:} Which step in Figure 1 maps to the content of Figure 10? \\
    \textbf{Evidence Pages:} [3, 14] \\
    \textbf{Ground Truth:} Deletion/duplication/rearrangement of the genetic material and Genetic diseases.
    \end{modelbox}

    \begin{modelbox}{green!10}{RAG$^{\text{txt+i2t}}_{\text{BGE-m3}}$ \xmark}
    \textbf{Response:}
    
    The question asks which step in \textbf{Figure~1} (Pathogenesis of genetic diseases) maps to the content of \textbf{Figure~10}. However, the provided context does not include any information about Figure~10. Since the content of Figure~10 is not described in the given text, it is impossible to determine which step in Figure~1 corresponds to it.
    
    \medskip
    \textbf{Final Answer: I don't know.}
    \end{modelbox}

    \begin{modelbox}{red!10}{VRAG$^{\text{page}}_{\text{ColPali}}$ \xmark}
    \textbf{Response:}
    
    The question asks which step in \textbf{Figure~1} maps to the content of \textbf{Figure~10}. However, the provided images and text do not include either Figure~1 or Figure~10. Without these figures, it is impossible to determine the mapping between them.
    
    \medskip
    \textbf{Final Answer: I don't know.}
    \end{modelbox}

    \begin{modelbox}{orange!15}{MLDocRAG \cmark}
    \textbf{Response:}
    
    To determine which step in Figure~1 maps to the content of Figure~10, let's analyze both figures step by step.
    
    \medskip
    \textbf{Figure~1: Pathogenesis of Genetic Diseases.}
    
    Figure~1 outlines a sequence of events leading to genetic diseases:
    \begin{enumerate}[leftmargin=*]
        \item \textbf{Mutation}
        \item \textbf{Deletion / Duplication / Rearrangement of Genetic Material}
        \item \textbf{Deficient / Defective Transcription of mRNA}
        \item \textbf{Deficient / Defective Post-Transcriptional Modifications of mRNA}
        \item \textbf{Deficient / Defective Translation of mRNA}
        \item \textbf{Deficient / Defective Production of Proteins / Regulatory Factors}
        \item \textbf{Deficient / Defective Post-Translational Modifications of Proteins}
        \item \textbf{Disturbed Function(s) of Metabolic Networks}
        \item \textbf{Patho-Physiological Alterations of Cellular Functions}
        \item \textbf{Genetic Disease}
    \end{enumerate}
    
    \medskip
    \textbf{Figure~10: Pathological Consequences of Genomic Structural and Functional Mutations.}
    
    Figure~10 illustrates the consequences of genomic imprinting defects and chromosomal abnormalities, specifically:
    \begin{itemize}
        \item \textbf{Structural and numerical chromosomal mutations} leading to altered genetic material.
        \item The resulting phenotypes (e.g., normal zygote, gynogenetic constitution, and androgenetic constitution) are shown.
    \end{itemize}
    
    \medskip
    \textbf{Mapping the Content.}
    
    Figure~10 focuses on the consequences of \textbf{structural and numerical chromosomal mutations}, which directly correspond to the second step in Figure~1:
    \textbf{Deletion / Duplication / Rearrangement of Genetic Material}.
    
    The phenotypes shown in Figure~10 (e.g., normal zygote, gynogenetic constitution, and androgenetic constitution) are the downstream effects of these chromosomal mutations, aligning with the pathogenesis outlined in Figure~1.
    
    \medskip
    \textbf{Final Answer: Deletion / Duplication / Rearrangement of Genetic Material.}
    \end{modelbox}

\end{minipage}

\vspace{-1.0\baselineskip}
\caption{Use Case on MMLongBench-Doc.}
\label{fig:case_study_layout}
\vspace{-1.0\baselineskip}
\end{figure}

\subsection{Use Case}~\label{subsect:use case}

Figure~\ref{fig:case_study_layout} illustrates a representative use case from MMLongBench-Doc that requires cross-modal reasoning across distant scientific diagrams (Pages~3 and~14).
In contrast to baseline methods such as $RAG^{txt+i2t}_{\text{BGE-m3}}$ and $VRAG^{page}_{\text{ColPali}}$, which fail to bridge this long-context gap, MLDocRAG successfully retrieves the dispersed evidence and performs the required multi-hop reasoning.
Specifically, MLDocRAG accurately aligns the ``chromosomal mutations'' visually depicted in ``Figure 10'' on Page~14 with the corresponding ``Deletion / Duplication / Rearrangement'' stage in ``Figure 1'' on Page~3, demonstrating its superior capability in fine-grained information extraction and cross-page visual alignment.
More use cases are provided in Appendix~\ref{use case: mldocrag qa example}.

\section{Conclusion}

We presented MLDocRAG, a framework for multimodal long-document QA built on the Multimodal Chunk-Query Graph (MCQG), which enables unified, query-centric retrieval.
By extending document expansion to the multimodal setting via MDoc2Query, MCQG organizes heterogeneous chunks and their generated queries into a structured graph that captures cross-modal and cross-page associations.
This query-centric representation supports selective, multi-hop retrieval and semantically grounded evidence aggregation, achieving consistent gains on MMLongBench-Doc and LongDocURL.
Our results demonstrate the effectiveness of query-based multimodal retrieval and the scalability of graph-structured organization for multimodal long-context understanding.

\section{Limitations}

While MLDocRAG achieves promising results, several limitations remain.  
First, it currently supports only text and image, limiting generalization to richer modalities such as video or audio.  
Second, its effectiveness depends on the quality of generated queries—noisy or incomplete queries may reduce retrieval accuracy.  
Finally, constructing large multimodal graphs can be computationally expensive, posing challenges for scaling to massive document collections.

\bibliographystyle{ACM-Reference-Format}
\bibliography{paper}

@article{nogueira2019document,
  title={Document expansion by query prediction},
  author={Nogueira, Rodrigo and Yang, Wei and Lin, Jimmy and Cho, Kyunghyun},
  journal={arXiv preprint arXiv:1904.08375},
  year={2019}
}

@article{wu2025query,
  title={Query-Centric Graph Retrieval Augmented Generation},
  author={Wu, Yaxiong and Bo, Jianyuan and Zhang, Yongyue and Liang, Sheng and Liu, Yong},
  journal={arXiv preprint arXiv:2509.21237},
  year={2025}
}

@article{hurst2024gpt,
  title={Gpt-4o system card},
  author={Hurst, Aaron and Lerer, Adam and Goucher, Adam P and Perelman, Adam and Ramesh, Aditya and Clark, Aidan and Ostrow, AJ and Welihinda, Akila and Hayes, Alan and Radford, Alec and others},
  journal={arXiv preprint arXiv:2410.21276},
  year={2024}
}

@article{team2023gemini,
  title={Gemini: a family of highly capable multimodal models},
  author={Team, Gemini and Anil, Rohan and Borgeaud, Sebastian and Alayrac, Jean-Baptiste and Yu, Jiahui and Soricut, Radu and Schalkwyk, Johan and Dai, Andrew M and Hauth, Anja and Millican, Katie and others},
  journal={arXiv preprint arXiv:2312.11805},
  year={2023}
}

@article{bai2025qwen2,
  title={Qwen2. 5-vl technical report},
  author={Bai, Shuai and Chen, Keqin and Liu, Xuejing and Wang, Jialin and Ge, Wenbin and Song, Sibo and Dang, Kai and Wang, Peng and Wang, Shijie and Tang, Jun and others},
  journal={arXiv preprint arXiv:2502.13923},
  year={2025}
}

@inproceedings{chen2024internvl,
  title={Internvl: Scaling up vision foundation models and aligning for generic visual-linguistic tasks},
  author={Chen, Zhe and Wu, Jiannan and Wang, Wenhai and Su, Weijie and Chen, Guo and Xing, Sen and Zhong, Muyan and Zhang, Qinglong and Zhu, Xizhou and Lu, Lewei and others},
  booktitle={Proceedings of the IEEE/CVF conference on computer vision and pattern recognition},
  pages={24185--24198},
  year={2024}
}

@inproceedings{wang2025multimodal,
  title={Multimodal needle in a haystack: Benchmarking long-context capability of multimodal large language models},
  author={Wang, Hengyi and Shi, Haizhou and Tan, Shiwei and Qin, Weiyi and Wang, Wenyuan and Zhang, Tunyu and Nambi, Akshay and Ganu, Tanuja and Wang, Hao},
  booktitle={Proceedings of the 2025 Conference of the Nations of the Americas Chapter of the Association for Computational Linguistics: Human Language Technologies (Volume 1: Long Papers)},
  pages={3221--3241},
  year={2025}
}

@article{zhao2023retrieving,
  title={Retrieving multimodal information for augmented generation: A survey},
  author={Zhao, Ruochen and Chen, Hailin and Wang, Weishi and Jiao, Fangkai and Do, Xuan Long and Qin, Chengwei and Ding, Bosheng and Guo, Xiaobao and Li, Minzhi and Li, Xingxuan and others},
  journal={arXiv preprint arXiv:2303.10868},
  year={2023}
}

@article{abootorabi2025ask,
  title={Ask in any modality: A comprehensive survey on multimodal retrieval-augmented generation},
  author={Abootorabi, Mohammad Mahdi and Zobeiri, Amirhosein and Dehghani, Mahdi and Mohammadkhani, Mohammadali and Mohammadi, Bardia and Ghahroodi, Omid and Baghshah, Mahdieh Soleymani and Asgari, Ehsaneddin},
  journal={arXiv preprint arXiv:2502.08826},
  year={2025}
}

@article{mei2025survey,
  title={A survey of multimodal retrieval-augmented generation},
  author={Mei, Lang and Mo, Siyu and Yang, Zhihan and Chen, Chong},
  journal={arXiv preprint arXiv:2504.08748},
  year={2025}
}

@inproceedings{radford2021learning,
  title={Learning transferable visual models from natural language supervision},
  author={Radford, Alec and Kim, Jong Wook and Hallacy, Chris and Ramesh, Aditya and Goh, Gabriel and Agarwal, Sandhini and Sastry, Girish and Askell, Amanda and Mishkin, Pamela and Clark, Jack and others},
  booktitle={International conference on machine learning},
  pages={8748--8763},
  year={2021},
  organization={PmLR}
}

@article{tschannen2025siglip,
  title={Siglip 2: Multilingual vision-language encoders with improved semantic understanding, localization, and dense features},
  author={Tschannen, Michael and Gritsenko, Alexey and Wang, Xiao and Naeem, Muhammad Ferjad and Alabdulmohsin, Ibrahim and Parthasarathy, Nikhil and Evans, Talfan and Beyer, Lucas and Xia, Ye and Mustafa, Basil and others},
  journal={arXiv preprint arXiv:2502.14786},
  year={2025}
}

@inproceedings{zhai2023sigmoid,
  title={Sigmoid loss for language image pre-training},
  author={Zhai, Xiaohua and Mustafa, Basil and Kolesnikov, Alexander and Beyer, Lucas},
  booktitle={Proceedings of the IEEE/CVF international conference on computer vision},
  pages={11975--11986},
  year={2023}
}

@article{faysse2024colpali,
  title={Colpali: Efficient document retrieval with vision language models},
  author={Faysse, Manuel and Sibille, Hugues and Wu, Tony and Omrani, Bilel and Viaud, Gautier and Hudelot, C{\'e}line and Colombo, Pierre},
  journal={arXiv preprint arXiv:2407.01449},
  year={2024}
}

@article{edge2024local,
  title={From local to global: A graph rag approach to query-focused summarization},
  author={Edge, Darren and Trinh, Ha and Cheng, Newman and Bradley, Joshua and Chao, Alex and Mody, Apurva and Truitt, Steven and Metropolitansky, Dasha and Ness, Robert Osazuwa and Larson, Jonathan},
  journal={arXiv preprint arXiv:2404.16130},
  year={2024}
}

@article{zhang2025comprehensive,
  title={A Comprehensive Survey on Multimodal RAG: All Combinations of Modalities as Input and Output},
  author={Zhang, Rui and Liu, Chen and Su, Yixin and Li, Ruixuan and Huang, Xuanjing and Li, Xuelong and Yu, Philip S},
  journal={Authorea Preprints},
  year={2025},
  publisher={Authorea}
}

@article{ma2024mmlongbench,
  title={Mmlongbench-doc: Benchmarking long-context document understanding with visualizations},
  author={Ma, Yubo and Zang, Yuhang and Chen, Liangyu and Chen, Meiqi and Jiao, Yizhu and Li, Xinze and Lu, Xinyuan and Liu, Ziyu and Ma, Yan and Dong, Xiaoyi and others},
  journal={Advances in Neural Information Processing Systems},
  volume={37},
  pages={95963--96010},
  year={2024}
}

@inproceedings{deng2025longdocurl,
  title={Longdocurl: a comprehensive multimodal long document benchmark integrating understanding, reasoning, and locating},
  author={Deng, Chao and Yuan, Jiale and Bu, Pi and Wang, Peijie and Li, Zhong-Zhi and Xu, Jian and Li, Xiao-Hui and Gao, Yuan and Song, Jun and Zheng, Bo and others},
  booktitle={Proceedings of the 63rd Annual Meeting of the Association for Computational Linguistics (Volume 1: Long Papers)},
  pages={1135--1159},
  year={2025}
}

@inproceedings{gospodinov2023doc2query,
  title={Doc2Query--: when less is more},
  author={Gospodinov, Mitko and MacAvaney, Sean and Macdonald, Craig},
  booktitle={European Conference on Information Retrieval},
  pages={414--422},
  year={2023},
  organization={Springer}
}

@article{cho2024m3docrag,
  title={M3docrag: Multi-modal retrieval is what you need for multi-page multi-document understanding},
  author={Cho, Jaemin and Mahata, Debanjan and Irsoy, Ozan and He, Yujie and Bansal, Mohit},
  journal={arXiv preprint arXiv:2411.04952},
  year={2024}
}

@inproceedings{van2023document,
  title={Document understanding dataset and evaluation (dude)},
  author={Van Landeghem, Jordy and Tito, Rub{\`e}n and Borchmann, {\L}ukasz and Pietruszka, Micha{\l} and Joziak, Pawel and Powalski, Rafal and Jurkiewicz, Dawid and Coustaty, Micka{\"e}l and Anckaert, Bertrand and Valveny, Ernest and others},
  booktitle={Proceedings of the IEEE/CVF International Conference on Computer Vision},
  pages={19528--19540},
  year={2023}
}

@article{park2025dochop,
  title={DocHop-QA: Towards Multi-Hop Reasoning over Multimodal Document Collections},
  author={Park, Jiwon and Pyeon, Seohyun and Kim, Jinwoo and Cabal, Rina Carines and Ding, Yihao and Han, Soyeon Caren},
  journal={arXiv preprint arXiv:2508.15851},
  year={2025}
}

@inproceedings{tanaka2023slidevqa,
  title={Slidevqa: A dataset for document visual question answering on multiple images},
  author={Tanaka, Ryota and Nishida, Kyosuke and Nishida, Kosuke and Hasegawa, Taku and Saito, Itsumi and Saito, Kuniko},
  booktitle={Proceedings of the AAAI Conference on Artificial Intelligence},
  volume={37},
  number={11},
  pages={13636--13645},
  year={2023}
}

@article{islam2023financebench,
  title={Financebench: A new benchmark for financial question answering},
  author={Islam, Pranab and Kannappan, Anand and Kiela, Douwe and Qian, Rebecca and Scherrer, Nino and Vidgen, Bertie},
  journal={arXiv preprint arXiv:2311.11944},
  year={2023}
}

@inproceedings{hu2024mplug,
  title={mplug-docowl 1.5: Unified structure learning for ocr-free document understanding},
  author={Hu, Anwen and Xu, Haiyang and Ye, Jiabo and Yan, Ming and Zhang, Liang and Zhang, Bo and Zhang, Ji and Jin, Qin and Huang, Fei and Zhou, Jingren},
  booktitle={Findings of the Association for Computational Linguistics: EMNLP 2024},
  pages={3096--3120},
  year={2024}
}

@article{dong2024internlm,
  title={Internlm-xcomposer2-4khd: A pioneering large vision-language model handling resolutions from 336 pixels to 4k hd},
  author={Dong, Xiaoyi and Zhang, Pan and Zang, Yuhang and Cao, Yuhang and Wang, Bin and Ouyang, Linke and Zhang, Songyang and Duan, Haodong and Zhang, Wenwei and Li, Yining and others},
  journal={Advances in Neural Information Processing Systems},
  volume={37},
  pages={42566--42592},
  year={2024}
}

@article{wang2024mineru,
  title={Mineru: An open-source solution for precise document content extraction},
  author={Wang, Bin and Xu, Chao and Zhao, Xiaomeng and Ouyang, Linke and Wu, Fan and Zhao, Zhiyuan and Xu, Rui and Liu, Kaiwen and Qu, Yuan and Shang, Fukai and others},
  journal={arXiv preprint arXiv:2409.18839},
  year={2024}
}

@article{robertson2009probabilistic,
  title={The probabilistic relevance framework: BM25 and beyond},
  author={Robertson, Stephen and Zaragoza, Hugo and others},
  journal={Foundations and trends{\textregistered} in information retrieval},
  volume={3},
  number={4},
  pages={333--389},
  year={2009},
  publisher={Now Publishers, Inc.}
}

@article{chen2024bge,
  title={Bge m3-embedding: Multi-lingual, multi-functionality, multi-granularity text embeddings through self-knowledge distillation},
  author={Chen, Jianlv and Xiao, Shitao and Zhang, Peitian and Luo, Kun and Lian, Defu and Liu, Zheng},
  journal={arXiv preprint arXiv:2402.03216},
  volume={4},
  number={5},
  year={2024}
}

@article{li2024survey,
  title={A survey on benchmarks of multimodal large language models},
  author={Li, Jian and Lu, Weiheng and Fei, Hao and Luo, Meng and Dai, Ming and Xia, Min and Jin, Yizhang and Gan, Zhenye and Qi, Ding and Fu, Chaoyou and others},
  journal={arXiv preprint arXiv:2408.08632},
  year={2024}
}

@article{wang2025mcot,
  title={Multimodal chain-of-thought reasoning: A comprehensive survey},
  author={Wang, Yaoting and Wu, Shengqiong and Zhang, Yuecheng and Yan, Shuicheng and Liu, Ziwei and Luo, Jiebo and Fei, Hao},
  journal={arXiv preprint arXiv:2503.12605},
  year={2025}
}

@article{yin2024survey,
  title={A survey on multimodal large language models},
  author={Yin, Shukang and Fu, Chaoyou and Zhao, Sirui and Li, Ke and Sun, Xing and Xu, Tong and Chen, Enhong},
  journal={National Science Review},
  volume={11},
  number={12},
  pages={nwae403},
  year={2024},
  publisher={Oxford University Press}
}

@article{liu2025comprehensive,
  title={A comprehensive survey on long context language modeling},
  author={Liu, Jiaheng and Zhu, Dawei and Bai, Zhiqi and He, Yancheng and Liao, Huanxuan and Que, Haoran and Wang, Zekun and Zhang, Chenchen and Zhang, Ge and Zhang, Jiebin and others},
  journal={arXiv preprint arXiv:2503.17407},
  year={2025}
}

@article{wang2024needle,
  title={Needle in a multimodal haystack},
  author={Wang, Weiyun and Zhang, Shuibo and Ren, Yiming and Duan, Yuchen and Li, Tiantong and Liu, Shuo and Hu, Mengkang and Chen, Zhe and Zhang, Kaipeng and Lu, Lewei and others},
  journal={Advances in Neural Information Processing Systems},
  volume={37},
  pages={20540--20565},
  year={2024}
}

@inproceedings{zhang2025ocr,
  title={Ocr hinders rag: Evaluating the cascading impact of ocr on retrieval-augmented generation},
  author={Zhang, Junyuan and Zhang, Qintong and Wang, Bin and Ouyang, Linke and Wen, Zichen and Li, Ying and Chow, Ka-Ho and He, Conghui and Zhang, Wentao},
  booktitle={Proceedings of the IEEE/CVF International Conference on Computer Vision},
  pages={17443--17453},
  year={2025}
}

@inproceedings{smith2007overview,
  title={An overview of the Tesseract OCR engine},
  author={Smith, Ray},
  booktitle={Ninth international conference on document analysis and recognition (ICDAR 2007)},
  volume={2},
  pages={629--633},
  year={2007},
  organization={IEEE}
}

@article{fan2025minirag,
  title={MiniRAG: Towards Extremely Simple Retrieval-Augmented Generation},
  author={Fan, Tianyu and Wang, Jingyuan and Ren, Xubin and Huang, Chao},
  journal={arXiv preprint arXiv:2501.06713},
  year={2025}
}

@article{hossain2019comprehensive,
  title={A comprehensive survey of deep learning for image captioning},
  author={Hossain, MD Zakir and Sohel, Ferdous and Shiratuddin, Mohd Fairuz and Laga, Hamid},
  journal={ACM Computing Surveys (CsUR)},
  volume={51},
  number={6},
  pages={1--36},
  year={2019},
  publisher={ACM New York, NY, USA}
}

@inproceedings{tao2006language,
  title={Language model information retrieval with document expansion},
  author={Tao, Tao and Wang, Xuanhui and Mei, Qiaozhu and Zhai, ChengXiang},
  booktitle={Proceedings of the Human Language Technology Conference of the NAACL, Main Conference},
  pages={407--414},
  year={2006}
}

@inproceedings{singhal1999document,
  title={Document expansion for speech retrieval},
  author={Singhal, Amit and Pereira, Fernando},
  booktitle={Proceedings of the 22nd annual international ACM SIGIR conference on Research and development in information retrieval},
  pages={34--41},
  year={1999}
}

@inproceedings{efron2012improving,
  title={Improving retrieval of short texts through document expansion},
  author={Efron, Miles and Organisciak, Peter and Fenlon, Katrina},
  booktitle={Proceedings of the 35th international ACM SIGIR conference on Research and development in information retrieval},
  pages={911--920},
  year={2012}
}

@inproceedings{wan2007single,
  title={Single document summarization with document expansion},
  author={Wan, Xiaojun and Yang, Jianwu and Xiao, Jianguo},
  booktitle={AAAI},
  pages={931--936},
  year={2007}
}

@article{aumuller2020ann,
  title={ANN-Benchmarks: A benchmarking tool for approximate nearest neighbor algorithms},
  author={Aum{\"u}ller, Martin and Bernhardsson, Erik and Faithfull, Alexander},
  journal={Information Systems},
  volume={87},
  pages={101374},
  year={2020},
  publisher={Elsevier}
}

@article{douze2025faiss,
  title={The faiss library},
  author={Douze, Matthijs and Guzhva, Alexandr and Deng, Chengqi and Johnson, Jeff and Szilvasy, Gergely and Mazar{\'e}, Pierre-Emmanuel and Lomeli, Maria and Hosseini, Lucas and J{\'e}gou, Herv{\'e}},
  journal={IEEE Transactions on Big Data},
  year={2025},
  publisher={IEEE}
}

@book{kuc2013elasticsearch,
  title={Elasticsearch server},
  author={Kuc, Rafal and Rogozinski, Marek},
  year={2013},
  publisher={Packt Publishing Ltd}
}

@inproceedings{guia2017graph,
  title={Graph databases: Neo4j analysis.},
  author={Guia, Jos{\'e} and Soares, Val{\'e}ria Gon{\c{c}}alves and Bernardino, Jorge},
  booktitle={ICEIS (1)},
  pages={351--356},
  year={2017}
}

@article{gu2024survey,
  title={A survey on llm-as-a-judge},
  author={Gu, Jiawei and Jiang, Xuhui and Shi, Zhichao and Tan, Hexiang and Zhai, Xuehao and Xu, Chengjin and Li, Wei and Shen, Yinghan and Ma, Shengjie and Liu, Honghao and others},
  journal={The Innovation},
  year={2024},
  publisher={Elsevier}
}

@misc{qwen2.5,
    title = {Qwen2.5: A Party of Foundation Models},
    url = {https://qwenlm.github.io/blog/qwen2.5/},
    author = {Qwen Team},
    month = {September},
    year = {2024}
}

@article{qwen2,
      title={Qwen2 Technical Report}, 
      author={An Yang and Baosong Yang and Binyuan Hui and Bo Zheng and Bowen Yu and Chang Zhou and Chengpeng Li and Chengyuan Li and Dayiheng Liu and Fei Huang and Guanting Dong and Haoran Wei and Huan Lin and Jialong Tang and Jialin Wang and Jian Yang and Jianhong Tu and Jianwei Zhang and Jianxin Ma and Jin Xu and Jingren Zhou and Jinze Bai and Jinzheng He and Junyang Lin and Kai Dang and Keming Lu and Keqin Chen and Kexin Yang and Mei Li and Mingfeng Xue and Na Ni and Pei Zhang and Peng Wang and Ru Peng and Rui Men and Ruize Gao and Runji Lin and Shijie Wang and Shuai Bai and Sinan Tan and Tianhang Zhu and Tianhao Li and Tianyu Liu and Wenbin Ge and Xiaodong Deng and Xiaohuan Zhou and Xingzhang Ren and Xinyu Zhang and Xipin Wei and Xuancheng Ren and Yang Fan and Yang Yao and Yichang Zhang and Yu Wan and Yunfei Chu and Yuqiong Liu and Zeyu Cui and Zhenru Zhang and Zhihao Fan},
      journal={arXiv preprint arXiv:2407.10671},
      year={2024}
}

@article{bai2025qwen2.5-vl,
  title={Qwen2.5-vl technical report},
  author={Bai, Shuai and Chen, Keqin and Liu, Xuejing and Wang, Jialin and Ge, Wenbin and Song, Sibo and Dang, Kai and Wang, Peng and Wang, Shijie and Tang, Jun and others},
  journal={arXiv preprint arXiv:2502.13923},
  year={2025}
}

@article{zheng2024sglang,
  title={Sglang: Efficient execution of structured language model programs},
  author={Zheng, Lianmin and Yin, Liangsheng and Xie, Zhiqiang and Sun, Chuyue Livia and Huang, Jeff and Yu, Cody Hao and Cao, Shiyi and Kozyrakis, Christos and Stoica, Ion and Gonzalez, Joseph E and others},
  journal={Advances in neural information processing systems},
  volume={37},
  pages={62557--62583},
  year={2024}
}

\appendix

\balance

\clearpage

\nobalance

\raggedbottom

\section{LLM Prompt}~\label{appendix:llm_prompt}
The following are some LLM prompts for query generation and answer evaluation, including MDoc2Query Prompt, Page-Context-Aware MDoc2Query Prompt, Response Generation Prompt, Page-Context-Aware Response Generation and Evaluation Prompt.

\subsection{MDoc2Query Prompt}

\begin{tcolorbox}[
    enhanced,
    breakable,
    colback=gray!5!white, 
    colframe=black!75!black,
    colbacktitle=black!60!white,
    title=MDoc2Query Prompt,
    fonttitle=\bfseries\large,
    width=\linewidth,
    arc=0mm,
    boxrule=1pt,
]
\footnotesize

\textbf{---Role---}

You are a \textbf{Doc2Query} assistant.

\vspace{0.5em}
\textbf{---Goal---}

Given a text chunk and image, generate no more than 20 distinct user queries that can be directly answered by that provided text chunk and image. For each query, also provide an exact answer and a relevance score.

\vspace{0.5em}
\textbf{---Generation Rules---}
\begin{enumerate}
    \item \textbf{Answerability:} Every query must be answerable using \textit{only} information in the text chunk and image.
    
    \item \textbf{Comprehensive coverage:} Collectively, all generated queries should cover all key ideas in the text chunk and image from different viewpoints or levels of detail.
    
    \item \textbf{Adaptive quantity:} Adjust the number of generated queries (0-20) according to the semantic richness and information value of both text and image.
    
    \item \textbf{Diversity requirements:} Ensure diversity along the following dimensions:
    \begin{itemize}
        \item \textit{Question-style variety:} Mix interrogative forms (who/what/why/how/where/when/did), imperative prompts ("List...", "Summarize..."), comparative questions, conditional or speculative forms, etc.
        \item \textit{Content-perspective variety:} Include queries on facts, definitions, methods, reasons, outcomes, examples, comparisons, limitations, and so on.
        \item \textit{Information granularity:} Combine macro (overall purpose, high-level summary) and micro (specific figures, terms, steps) queries.
        \item \textit{User-intent variety:} Simulate intents such as confirmation, evaluation, usability, diagnosis, and decision-making (e.g., "Is this approach more efficient than...?").
        \item \textit{Linguistic expression variety:} Vary wording, syntax (active $\leftrightarrow$ passive), and synonyms; avoid repeating near-identical phrases.
        \item \textit{No redundancy:} Each query must be meaningfully distinct; eliminate trivial rephrases that offer no new angle.
        \item \textit{Chunk-grounded specificity:} Queries must be grounded in specific factual points from the text chunk and image. Avoid vague or generic formulations such as "What did X say?" or "Tell me more about Y" that lack anchoring in actual content.
    \end{itemize}

    \item \textbf{Required fields:} Each output item must be based on the given text chunk and image, including the following fields:
    \begin{itemize}
        \item \texttt{query}: A question or search phrase a user might ask.
        \item \texttt{answer}: A concise answer taken verbatim (or nearly verbatim) from the text chunk and image.
    \end{itemize}
\end{enumerate}

\vspace{0.5em}
\textbf{---Example---}

\textbf{1. Input Chunk and Images}
\begin{lstlisting}[style=jsonstyle, basicstyle=\ttfamily\scriptsize]
[
  {"type": "images", "image": "/path/image1"}, 
  {"type": "images", "image": "/path/image2"}, 
  {"type": "text", "text": "Alice met with Bob at the Central Cafe on Tuesday to discuss their upcoming collaborative research project..."}
]
\end{lstlisting}

\textbf{2. Generated Queries}
\begin{itemize}
    \item Where did Alice and Bob meet?
    \item When did the meeting take place?
    \item What was the main topic discussed during the meeting?
    \item Who suggested incorporating advanced AI methodologies?
    \item ... (more queries)
\end{itemize}

\vspace{0.5em}
\textbf{---Output Format---}

Return only a JSON array of objects. Each object must include:
\begin{itemize}
    \item \texttt{"index"}: a zero-based integer
    \item \texttt{"query"}: the generated question
    \item \texttt{"answer"}: the exact answer
\end{itemize}
Adjust the number of queries according to the information richness of the text and image (0–20 queries).

\begin{lstlisting}[style=jsonstyle]
[
  {"index": 0, "query": "Where did Alice and Bob meet?", "answer": "Central Cafe"},
  {"index": 1, "query": "When did the meeting take place?", "answer": "Tuesday"}
]
\end{lstlisting}

\end{tcolorbox}
\label{llm prompt:mdoc2query prompt}

\subsection{Page-Context-Aware MDoc2Query Prompt}

\begin{tcolorbox}[
    enhanced,
    breakable,               
    colback=gray!5!white, 
    colframe=black!75!black, 
    colbacktitle=black!60!white,
    title=Page-Context-Aware MDoc2Query Prompt,
    fonttitle=\bfseries\large,
    width=\linewidth,
    arc=0mm, 
    boxrule=1pt,
]
\footnotesize

\textbf{---Role---}

You are an expert **Document Understanding AI** designed to build a high-precision Retrieval Graph.

\vspace{0.5em}
\textbf{---Inputs---}
\begin{enumerate}
    \item \textbf{Target Chunk:} A specific segment of content (Text, Cropped Image, or Table Data) extracted from a document.
    \item \textbf{Source Page Image:} The original full-page screenshot where this chunk is located.
\end{enumerate}

\vspace{0.5em}
\textbf{---Workflow (Strict Execution Order)---}

\textbf{Step 1: Visual Localization \& Context Recovery (Mental Scratchpad)}\\
Locate: First, look at the Source Page Image and identify exactly where the Target Chunk is located.\\
Scan Surroundings: Look immediately above, below, and to the sides of the chunk location in the full page.\\
Identify Missing Context: Find any information \textit{not} inside the chunk but essential for understanding it. This includes:
\begin{itemize}
    \item Headers: Section Titles, Page Headers, Chapter Names.
    \item Captions: Figure Titles (e.g., "Figure 3: Revenue Trend"), Table Headers, or Axis Labels that were cut off.
    \item Pre-text: Preceding sentences that define what "this table" or "the data below" refers to.
\end{itemize}

\textbf{Step 2: Adaptive Query Generation}\\
Based on the chunk's content and the recovered context from Step 1, generate a set of 5 to 20 Query-Answer pairs.\\
\textit{Quantity Rule:} If the chunk is dense (e.g., a complex table or dense text), generate closer to 20. If it is sparse, generate closer to 5.\\
\textit{Coverage Rule:} You must generate queries for all 4 Levels defined below. The distribution ratio is up to you based on the content type.

\vspace{0.5em}
\textbf{---Query Level Definitions---}

\textbf{Level 1: Integrated Entity Relationships (Dense \& Comprehensive)}\\
Goal: Instead of asking about single entities one by one, generate complex queries that link multiple entities found in the chunk.\\
Constraint: Do not generate simple questions like "What is X?". Instead, ask: "How does [Entity A] interact with [Entity B] regarding [Topic]?" or "What are the key specifications and performance metrics of [Product X]?"\\
Why: To create strong semantic connections between entities in the graph without flooding it with simple queries.

\textbf{Level 2: Detailed Content Description (The "Core Message")}\\
Goal: Paraphrase and summarize the detailed information provided inside the chunk.\\
Instruction: Imagine a user asks "What specific details does this paragraph/chart provide?". Cover the key arguments, data trends, or descriptive points.\\
Why: To ensure the chunk is retrievable via natural language descriptions of its content.

\textbf{Level 3: Macro Hierarchy (Navigation)}\\
Goal: Anchor this chunk to the document structure.\\
Instruction: Extract the Section Title, Chapter Name, or Page Header from the Source Page Image. Generate queries that link this specific chunk content to that high-level topic.\\
Example: "What information does the section '[Section Title]' provide regarding [Chunk Topic]?"

\textbf{Level 4: Context Restoration (Immediate Surroundings)}\\
Goal: Fix "context loss" caused by chunking.\\
Instruction: Explicitly incorporate the missing headers, captions, or pre-text you found in Step 1 into the query.\\
Critical: If the chunk is a table/chart without a title, use the title found in the Page Image to formulate the query.\\
Example: "According to [Figure Title found in Page Image], what trend is shown in the data regarding [Chunk Content]?" or "As detailed in [Table Caption], what are the specific values for [Row Name]?"

\vspace{0.5em}
\textbf{---Output Format (JSON Only)---}

\begin{lstlisting}[style=jsonstyle]
[
    {
      "index": 0,
      "level": "level_1_entity_integrated",
      "query": "Comprehensive query connecting Entity A and B...",
      "answer": "Detailed answer..."
    },
    {
      "index": 1,
      "level": "level_2_detailed_content",
      "query": "...",
      "answer": "..."
    },
    {
      "index": 2,
      "level": "level_3_macro_hierarchy",
      "query": "...",
      "answer": "..."
    },
    {
      "index": 3,
      "level": "level_4_context_restoration",
      "query": "Query incorporating the external Figure Title/Table Header...",
      "answer": "..."
    }
    // ... generate 5 to 20 pairs total
]
\end{lstlisting}

\end{tcolorbox}
\label{llm prompt: page-context mdoc2query prompt}

\subsection{Response Generation Prompt}

\begin{tcolorbox}[
    enhanced,
    breakable,
    colback=gray!5!white, 
    colframe=black!75!black, 
    colbacktitle=black!60!white,
    title=Response Generation Prompt, 
    fonttitle=\bfseries\large,
    width=\linewidth,
    arc=0mm,
    boxrule=1pt,
]
\footnotesize

You are a knowledgeable assistant that answers questions based on the given text and image data.

\vspace{0.5em}
\textbf{---Guidelines---}
\begin{enumerate}
    \item Carefully reason through the provided information before answering, but only use evidence \textbf{explicitly supported} by the text or image.
    \item If the answer cannot be determined from the provided data, clearly say you don't know.
    \item Avoid unnecessary explanations---respond concisely and directly.
    \item Present the final output in the format: \textbf{Final Answer: [your answer]}
\end{enumerate}

\end{tcolorbox}
\label{llm prompt:response generation prompt}

\subsection{Page-Context-Aware Response Generation Prompt}

\begin{tcolorbox}[
    enhanced,
    breakable,
    colback=gray!5!white, 
    colframe=black!75!black, 
    colbacktitle=black!60!white,
    title=Page-Context-Aware Response Generation Prompt,
    fonttitle=\bfseries\large,
    width=\linewidth,
    arc=0mm,
    boxrule=1pt,
]
\footnotesize

You are an expert Multimodal QA Assistant. You will be provided with a user question and a set of \textbf{Retrieved Contexts}.
Each Context consists of:
\begin{enumerate}
    \item \textbf{Text/Data Chunk:} A specific segment of text, a table row, or a data point retrieved from a document.
    \item \textbf{Reference Page Image:} The full document page where this chunk is located.
\end{enumerate}

\vspace{0.5em}
\textbf{---Goal---}

Answer the user's question accurately by synthesizing information from the provided Text Chunks and verifying it against the Reference Page Images.

\vspace{0.5em}
\textbf{---Reasoning Guidelines---}

\begin{enumerate}
    \item \textbf{Visual Verification (Grounding):}
    \begin{itemize}
        \item Use the \textbf{Reference Page Image} to verify the context of the \textbf{Text Chunk}.
        \item \textit{Example:} If a chunk is a row of numbers, look at the Page Image to identify the \textbf{Column Headers} and \textbf{Row Labels} to ensure you interpret the numbers correctly.
        \item \textit{Example:} If a chunk describes a chart trend, look at the Page Image to confirm the \textbf{Axis Labels}, \textbf{Units}, and \textbf{Legends}.
    \end{itemize}

    \item \textbf{Contextual Synthesis:}
    \begin{itemize}
        \item The Text Chunk might be stripped of its section title. Use the Page Image to see which \textbf{Section Header} (e.g., "2023 Q4 Results" vs "2022 Q4 Results") the chunk belongs to.
        \item Combine information from multiple chunks if necessary to form a complete answer.
    \end{itemize}

    \item \textbf{Strict Evidence-Based:}
    \begin{itemize}
        \item Answer \textbf{ONLY} using the provided information. Do not use outside knowledge.
        \item If the text chunk and the image contradict each other, trust the \textbf{Visual Evidence (Image)} for raw data (numbers, charts) and the \textbf{Text Chunk} for semantic explanations.
        \item If the answer is not present in the provided contexts, explicitly state: "Based on the provided documents, I cannot answer this question."
    \end{itemize}

    \item \textbf{Conciseness:}
    \begin{itemize}
        \item Get straight to the point. Do not start with "Based on the context..." or "The image shows...". Just state the answer.
    \end{itemize}
\end{enumerate}

\vspace{0.5em}
\textbf{---Output Format---}

\begin{description}
    \item[Analysis:] (Briefly map the chunk text to the visual location in the page image to confirm headers/legends)
    \item[Final Answer:] [Your direct answer]
\end{description}

\end{tcolorbox}
\label{llm prompt:page-context response generation prompt}

\subsection{Evaluation Prompt}

\begin{tcolorbox}[
    enhanced,
    breakable,
    colback=gray!5!white, 
    colframe=black!75!black, 
    colbacktitle=black!60!white,
    title=Evaluation Prompt, 
    fonttitle=\bfseries\large,
    width=\linewidth,
    arc=0mm,
    boxrule=1pt,
]
\footnotesize

You are a strict and precise evaluation assistant. You will be given a question, a reference answer, and a candidate answer generated via retrieval-augmented generation (RAG).

\vspace{0.5em}
\textbf{---Goal---}

Evaluate the candidate answer against the reference answer based on factual accuracy and completeness. Slight differences in phrasing are acceptable as long as the meaning is the same.\\
If the candidate answer includes an analysis followed by "Final Answer:", only evaluate the content after "Final Answer:". If "Final Answer:" is missing, treat the entire candidate as the final answer.

\vspace{0.5em}
\textbf{---Normalization for Unanswerable/None---}

Treat the following expressions as \textbf{equivalent to "Not answerable / No answer / None"} when they appear as the candidate's final answer:
\begin{itemize}
    \item \textit{Explicit statements:} "I don't know", "Not answerable", "Not enough information", "Insufficient information", "Not mentioned", "Unknown", "Cannot be determined", "No information provided", "N/A", "Not applicable", "None".
    \item \textit{Negative-existence statements:} Assertions of absence for list/type questions, e.g., "No stages require a cooler.", "No such category exists.", "There are none.", "No [items] are present.", "None of the stages...".
\end{itemize}
\textbf{Note:} This normalization applies \textbf{only when the reference answer itself is unanswerable/none/empty}. Normalization \textbf{takes precedence over all other rules} when determining equivalence.

\vspace{0.5em}
\textbf{---Important Rules---}
\begin{itemize}
    \item If the candidate answer fails to provide an answer (e.g., says "I don't know") \textbf{when the reference is answerable}, it must receive a \textbf{score of 0}.
    \item If the reference answer is unanswerable/none/empty, the candidate answer must produce an unanswerable-equivalent expression (as normalized above) to receive \textbf{score 1}. If the candidate gives any substantive or fabricated specific content, assign \textbf{score 0}.
    \item If the candidate answer overgeneralizes, omits key elements, or adds unrelated information not supported by the reference, the score must be \textbf{0}, even if part of it is correct.
    \item Only when the candidate answer covers \textbf{all essential factual elements} of the reference answer without introducing unrelated content, should the score be \textbf{1}.
\end{itemize}

\vspace{0.5em}
\textbf{---Scoring Criteria---}

\textbf{Score = 1 (Correct):}
\begin{itemize}
    \item The candidate answer accurately matches the factual content and level of detail of the reference answer. Minor wording differences are acceptable if the meaning is equivalent.
    \item If the reference answer is "Unable to answer" / "Not answerable" / "None" / empty, and the candidate provides an unanswerable-equivalent expression (after normalization).
\end{itemize}

\textbf{Score = 0 (Incorrect):}
\begin{itemize}
    \item The reference answer is answerable, but the candidate says any unanswerable-equivalent expression (including negative-existence).
    \item The reference answer is answerable, but the candidate gives an incorrect, incomplete, or irrelevant answer.
    \item The reference answer is unanswerable/none/empty, but the candidate provides any substantive or fabricated answer.
\end{itemize}

\end{tcolorbox}
\label{llm prompt:evaluation prompt}

\section{Details of Visual Noise Filtering}
\label{appendix:visual_filtering}

We employ \texttt{clip-vit-base-patch32} to classify visual chunks via zero-shot inference. Based on their semantic relevance to the RAG task, visual elements are partitioned into two sets: those retained for indexing and those filtered as decorative noise. Table~\ref{tab:visual_categories} lists the specific labels for each action.

\begin{table}[ht]
\centering
\small
\begin{tabular}{l|p{0.75\columnwidth}}
\toprule
\textbf{Retain} & table, chart, graph, diagram, map, infographic, equation, flow chart, scatter plot, bar chart, form \\
\midrule
\textbf{Filter} & logo, banner, advertisement, poster, cover, illustration, background, icon, photo, texture \\
\bottomrule
\end{tabular}
\vspace{3mm}
\caption{Visual chunk classification and filtering actions.}
\label{tab:visual_categories}
\end{table}

\noindent \textbf{Rationale:} Elements labeled for filtering are pruned to eliminate redundant visual noise (e.g., repeating logos) and enhance the precision of the embedding space during retrieval.


\section{MDoc2Query Example}~\label{use case: mdoc2query example}
There are some query-answer pair examples generated by MDoc2Query and Page-Context-Aware MDoc2Query. The multimodal chunk and corresponding page image as follows:

\begin{tcolorbox}[
    enhanced,
    breakable,
    colback=gray!5!white, 
    colframe=black!75!black,   
    colbacktitle=black!60!white,
    title=Multimodal Chunk with Page Image,
    fonttitle=\bfseries\large,
    width=\linewidth,
    arc=0mm,
    boxrule=1pt,
]
\footnotesize

\begin{lstlisting}[style=jsonstyle]
{
    "type": "image",
    "text": "",
    "image": [
        "./images/9bc7afb9a74e615b67303104c415d.jpg"
        ],
    "visual_context": "./visual_context/a325667e4e534f96.png"
}
\end{lstlisting}

Chunk Image:
\par
\vspace{0.5em} 

{\centering
    \includegraphics[width=0.9\linewidth, keepaspectratio]{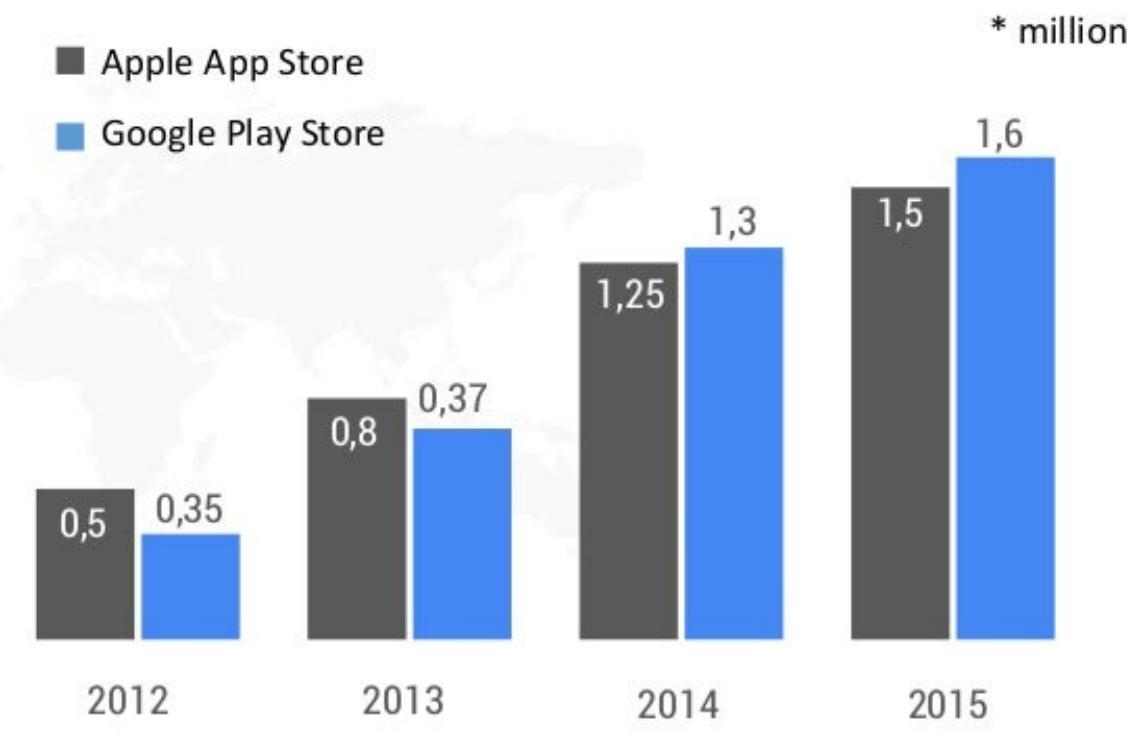}
\par}

Corresponding Page Image:
\par
\vspace{0.5em}

{\centering
    \includegraphics[width=0.9\linewidth, keepaspectratio]{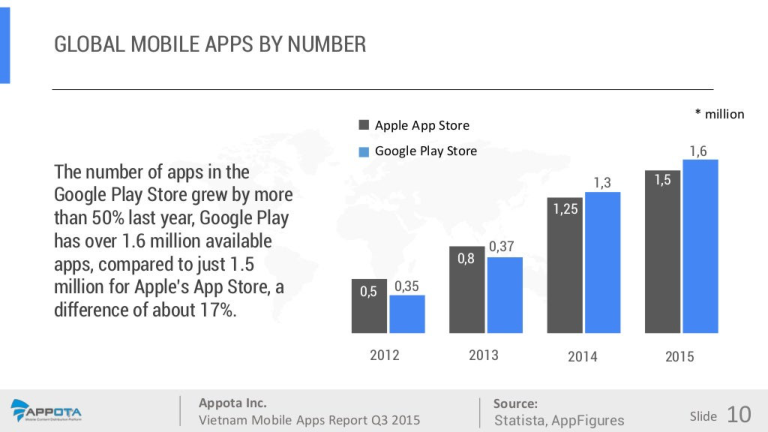}
\par}

\end{tcolorbox}
\label{llm prompt:response generation prompt}

\begin{tcolorbox}[
    enhanced,
    breakable,
    colback=gray!5!white,
    colframe=black!75!black,
    colbacktitle=black!60!white,
    title=Doc2Query Example,
    fonttitle=\bfseries\large,
    width=\linewidth,
    arc=0mm,
    boxrule=1pt,
]

\begin{lstlisting}[style=jsonstyle]
[
    {
        "index": 0,
        "query": "Which app store had more apps in 2015?",
        "answer": "Google Play Store",
        "q_id": "reportq32015-151009093138-lva1-app6891_95-6-0"
    },
    {
        "index": 1,
        "query": "How many apps were in the Apple App Store in 2015?",
        "answer": "1,5 million",
        "q_id": "reportq32015-151009093138-lva1-app6891_95-6-1"
    },
    {
        "index": 2,
        "query": "What was the number of apps in the Google Play Store in 2015?",
        "answer": "1,6 million",
        "q_id": "reportq32015-151009093138-lva1-app6891_95-6-2"
    },
    {
        "index": 3,
        "query": "In which year did the number of apps in both stores start to increase significantly?",
        "answer": "2013",
        "q_id": "reportq32015-151009093138-lva1-app6891_95-6-3"
    },
    {
        "index": 4,
        "query": "How many apps were in the Apple App Store in 2012?",
        "answer": "0,5 million",
        "q_id": "reportq32015-151009093138-lva1-app6891_95-6-4"
    },
    {
        "index": 5,
        "query": "What was the number of apps in the Google Play Store in 2012?",
        "answer": "0,35 million",
        "q_id": "reportq32015-151009093138-lva1-app6891_95-6-5"
    },
    ...
]
\end{lstlisting}

\end{tcolorbox}
\label{use case: doc2Query example}

\begin{tcolorbox}[
    enhanced,
    breakable,
    colback=gray!5!white,
    colframe=black!75!black,
    colbacktitle=black!60!white,
    title=Page-Context-Aware Example,
    fonttitle=\bfseries\large,
    width=\linewidth,
    arc=0mm,
    boxrule=1pt,
]

\begin{lstlisting}[style=jsonstyle]
[
    {
        "index": 0,
        "level": "level_1_entity_integrated",
        "query": "How do the growth trends of apps in the Apple App Store and Google Play Store compare from 2012 to 2015, and what does this indicate about their market dynamics?",
        "answer": "From 2012 to 2015, both the Apple App Store and Google Play Store experienced significant growth in the number of apps. The Google Play Store showed a more rapid increase, starting at 0.35 million in 2012 and reaching 1.6 million in 2015. The Apple App Store started at 0.5 million in 2012 and reached 1.5 million in 2015. This indicates that while both platforms grew, Google Play Store expanded at a faster rate, leading to a larger number of apps by 2015.",
        "q_id": "reportq32015-151009093138-lva1-app6891_95-6-0"
    },
    {
        "index": 1,
        "level": "level_2_detailed_content",
        "query": "What specific details does the chart provide about the number of apps in the Apple App Store and Google Play Store from 2012 to 2015?",
        "answer": "The chart shows the number of apps in the Apple App Store and Google Play Store from 2012 to 2015, measured in millions. In 2012, the Apple App Store had 0.5 million apps, while the Google Play Store had 0.35 million. By 2013, the numbers increased to 0.8 million for the Apple App Store and 0.37 million for the Google Play Store. In 2014, the Apple App Store reached 1.25 million apps, and the Google Play Store reached 1.3 million. By 2015, the Apple App Store had 1.5 million apps, and the Google Play Store had 1.6 million apps.",
        "q_id": "reportq32015-151009093138-lva1-app6891_95-6-1"
    },
    {
        "index": 2,
        "level": "level_3_macro_hierarchy",
        "query": "What information does the section 'Global Mobile Apps by Number' provide regarding the growth of apps in the Apple App Store and Google Play Store?",
        "answer": "The section 'Global Mobile Apps by Number' provides information on the growth of apps in the Apple App Store and Google Play Store from 2012 to 2015. It highlights that the number of apps in the Google Play Store grew by more than 50% last year, reaching over 1.6 million, compared to 1.5 million for the Apple App Store. This section emphasizes the rapid expansion of mobile apps globally and the competitive landscape between the two major app stores.",
        "q_id": "reportq32015-151009093138-lva1-app6891_95-6-2"
    },
    {
        "index": 3,
        "level": "level_4_context_restoration",
        "query": "According to the 'Global Mobile Apps by Number' section, what trend is shown in the data regarding the growth of apps in the Apple App Store and Google Play Store?",
        "answer": "The 'Global Mobile Apps by Number' section shows a consistent upward trend in the number of apps in both the Apple App Store and Google Play Store from 2012 to 2015. The Google Play Store experienced a more significant growth rate, surpassing the Apple App Store by 2015. This trend indicates the rapid expansion of the mobile app market and the increasing competition between the two major app stores.",
        "q_id": "reportq32015-151009093138-lva1-app6891_95-6-3"
    },
    ...
]
\end{lstlisting}

\end{tcolorbox}
\label{use case: page-context-aware doc2Query example}

\section{MLDocRAG QA Example}~\label{use case:QA}

There is a use case from the MMLongBench-Doc dataset to evaluate the multi-page and cross-modal reasoning capabilities on baseline method and our MLDocRAG.

\begin{tcolorbox}[
    enhanced,
    breakable,
    colback=gray!5!white,  
    colframe=black!75!black,  
    colbacktitle=black!60!white,
    title=QA Example 1 on MMLongBench-Doc, 
    fonttitle=\bfseries\large,
    width=\linewidth,
    arc=0mm,
    boxrule=1pt,
]
\footnotesize

Souce evidence images:
\par 
\vspace{0.5em} 

{\centering
    \includegraphics[width=1.0\linewidth, keepaspectratio]{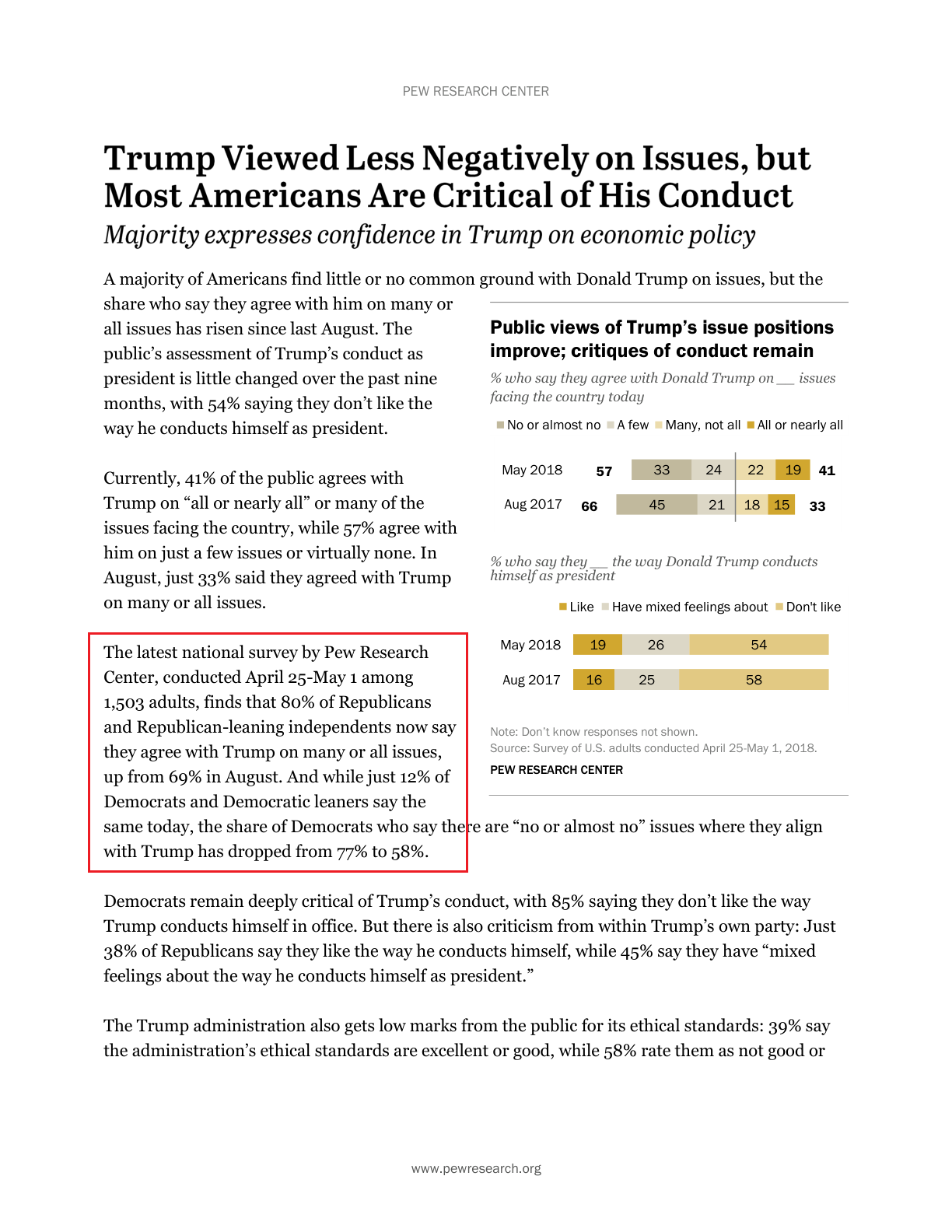}
\par}

\par 
\vspace{0.5em} 

{\centering
    \includegraphics[width=1.0\linewidth, keepaspectratio]{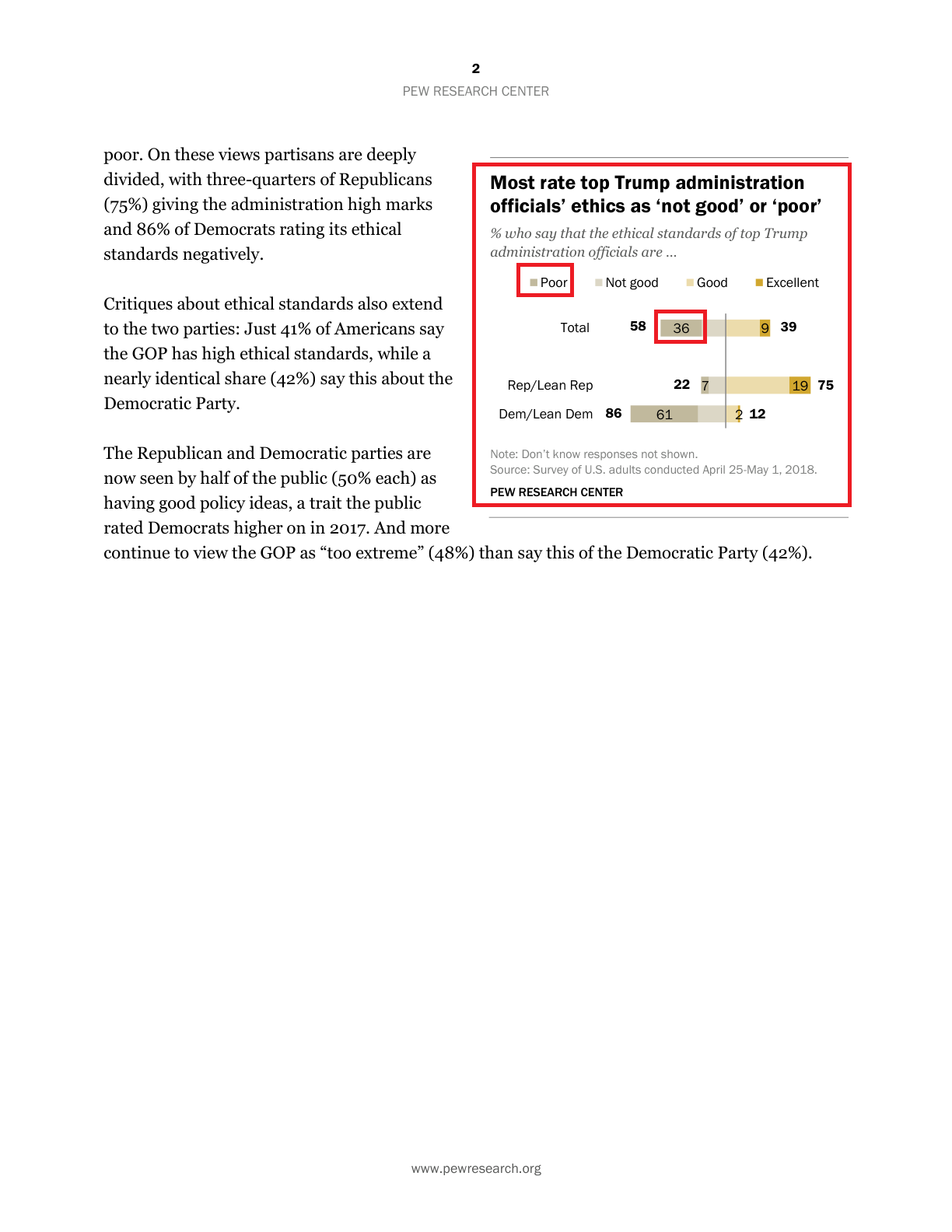}
\par}

\begin{minipage}{\linewidth}
    \textbf{Question:} Among the adults conducted the survey on April 25 - May 1 2018, how many adults rated Trump's government ethical standards as poor? \\
    \textbf{Evidence Pages:} [3, 4] \\
    \textbf{Ground Truth:} 541
\end{minipage}

\vspace{2mm} 

\begin{modelbox}{green!10}{RAG$^{\text{txt+i2t}}_{\text{BGE-m3}}$ \xmark}
\textbf{Response:}

To answer the question, let's break down the information provided:

\begin{enumerate}[leftmargin=*]
    \item \textbf{Overall Ratings of Trump's Government Ethical Standards:}
    \begin{itemize}
        \item The survey conducted from April~25 to May~1, 2018, asked respondents to rate the ethical standards of top Trump administration officials.
        \item According to the text, 58\% of adults rated the ethical standards of top Trump administration officials as ``not good'' or ``poor.''
    \end{itemize}

    \item \textbf{Specific Breakdown:}
    \begin{itemize}
        \item \textbf{Poor:} The text does not provide a specific percentage for those who rated the ethical standards as ``poor'' alone.
        \item However, it does state that 36\% of adults rated the ethical standards as ``poor.''
    \end{itemize}

    \item \textbf{Conclusion:}
    \begin{itemize}
        \item The question asks for the number of adults who rated Trump's government ethical standards as ``poor.''
        \item Based on the information provided, 36\% of adults rated the ethical standards as ``poor.''
    \end{itemize}
\end{enumerate}

\medskip
\textbf{Final Answer: 36\%}
\end{modelbox}

\vspace{1mm}

\begin{modelbox}{red!10}{VRAG$^{\text{page}}_{\text{ColPali}}$ \xmark}
\textbf{Response:}

To answer the question, let's analyze the relevant information from the provided images:

\begin{enumerate}[leftmargin=*]
    \item The survey was conducted from April~25 to May~1, 2018, as stated in the text.
    \item The relevant chart is the one titled ``Most rate top Trump administration officials' ethics as `not good' or `poor''' (Image~2).
    \item This chart shows the percentage of adults who rated the ethical standards of top Trump administration officials as ``Poor,'' ``Not good,'' ``Good,'' or ``Excellent.''
    \item The ``Poor'' category is represented by the dark blue bar, which shows 58\% of the total respondents.
\end{enumerate}

\medskip
\textbf{Final Answer: 58\% of adults rated Trump's government ethical standards as poor.}
\end{modelbox}

\vspace{1mm}

\begin{modelbox}{orange!15}{\texttt{MLDocRAG} \cmark}
\textbf{Response:}

To answer the question, let's break down the information provided:

\begin{enumerate}[leftmargin=*]
    \item The survey was conducted from April~25 to May~1, 2018.
    \item The question asks about the number of adults who rated Trump's government ethical standards as ``poor.''
    \item From the text, we know:
    \begin{itemize}
        \item About four-in-ten Americans rate the ethical standards of Trump administration officials as excellent (9\%) or good (30\%).
        \item 58\% say they are not good (21\%) or poor (36\%).
    \end{itemize}

    \item Specifically, the text states: ``58\% say they are not good (21\%) or poor (36\%).'' This means 36\% of the adults rated Trump's government ethical standards as ``poor.''
    \item The survey sample size is 1,503 adults.
    \item To find the number of adults who rated Trump's government ethical standards as ``poor,'' we calculate:
    \begin{itemize}
        \item 36\% of 1,503 = 0.36 $\times$ 1,503 = 541.08
    \end{itemize}
\end{enumerate}

Since we can't have a fraction of a person, we round to the nearest whole number.

\medskip
\textbf{Final Answer: 541}
\end{modelbox}

\end{tcolorbox}
\label{use case: mldocrag qa example}

Section~\ref{subsect:use case} presents a QA example on MMLongBench-Doc with detailed result comparisons; in this section, we further supplement the example with two evidence page images.

\begin{tcolorbox}[
    enhanced,
    breakable,
    colback=gray!5!white,  
    colframe=black!75!black,  
    colbacktitle=black!60!white,
    title=QA Example 2 on MMLongBench-Doc, 
    fonttitle=\bfseries\large,
    width=\linewidth,
    arc=0mm,
    boxrule=1pt,
]
\footnotesize

Souce evidence images:
\par 
\vspace{0.5em} 

{\centering
    \includegraphics[width=1.0\linewidth, keepaspectratio]{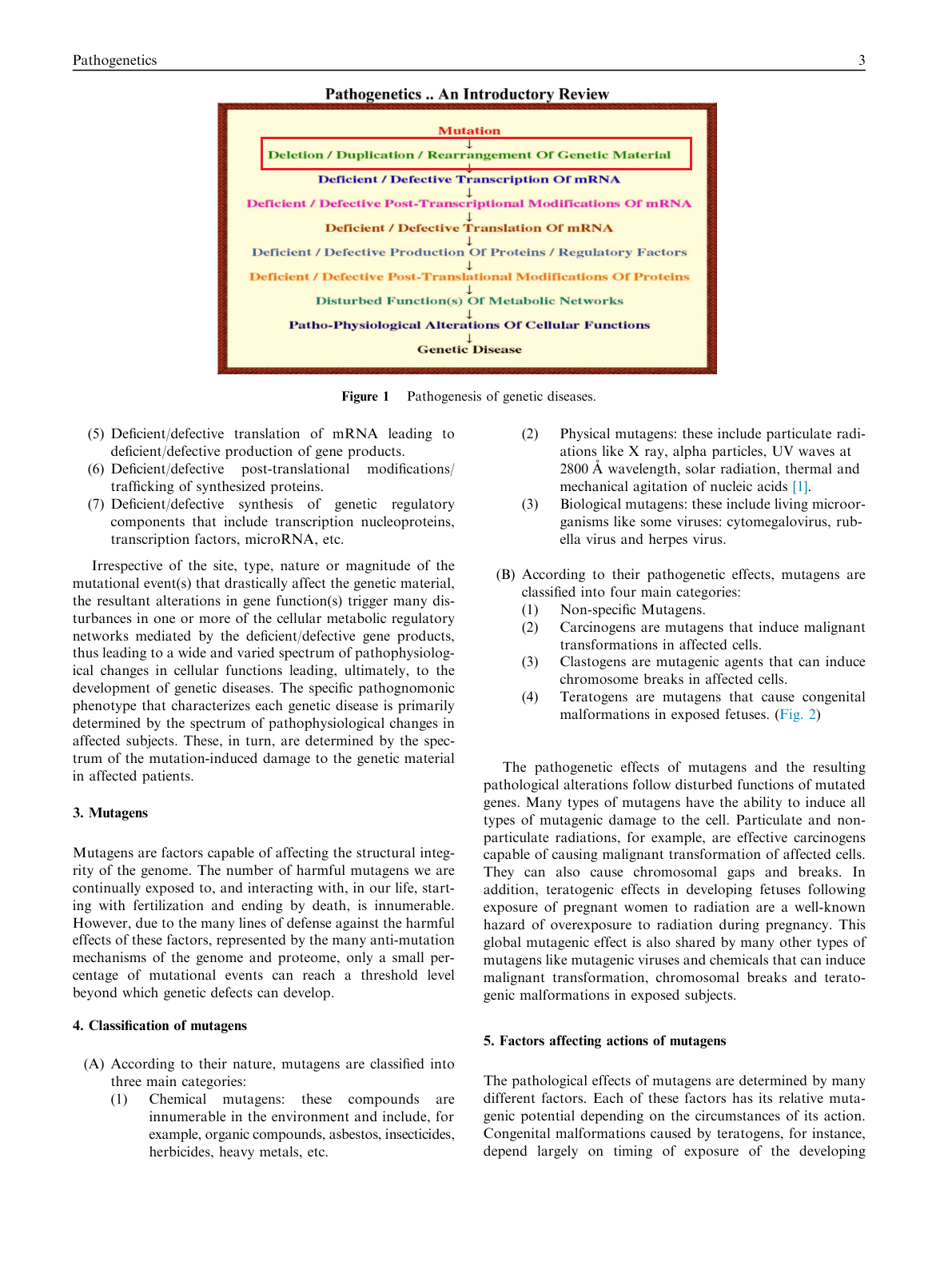}
\par}

\par 
\vspace{0.5em} 

{\centering
    \includegraphics[width=1.0\linewidth, keepaspectratio]{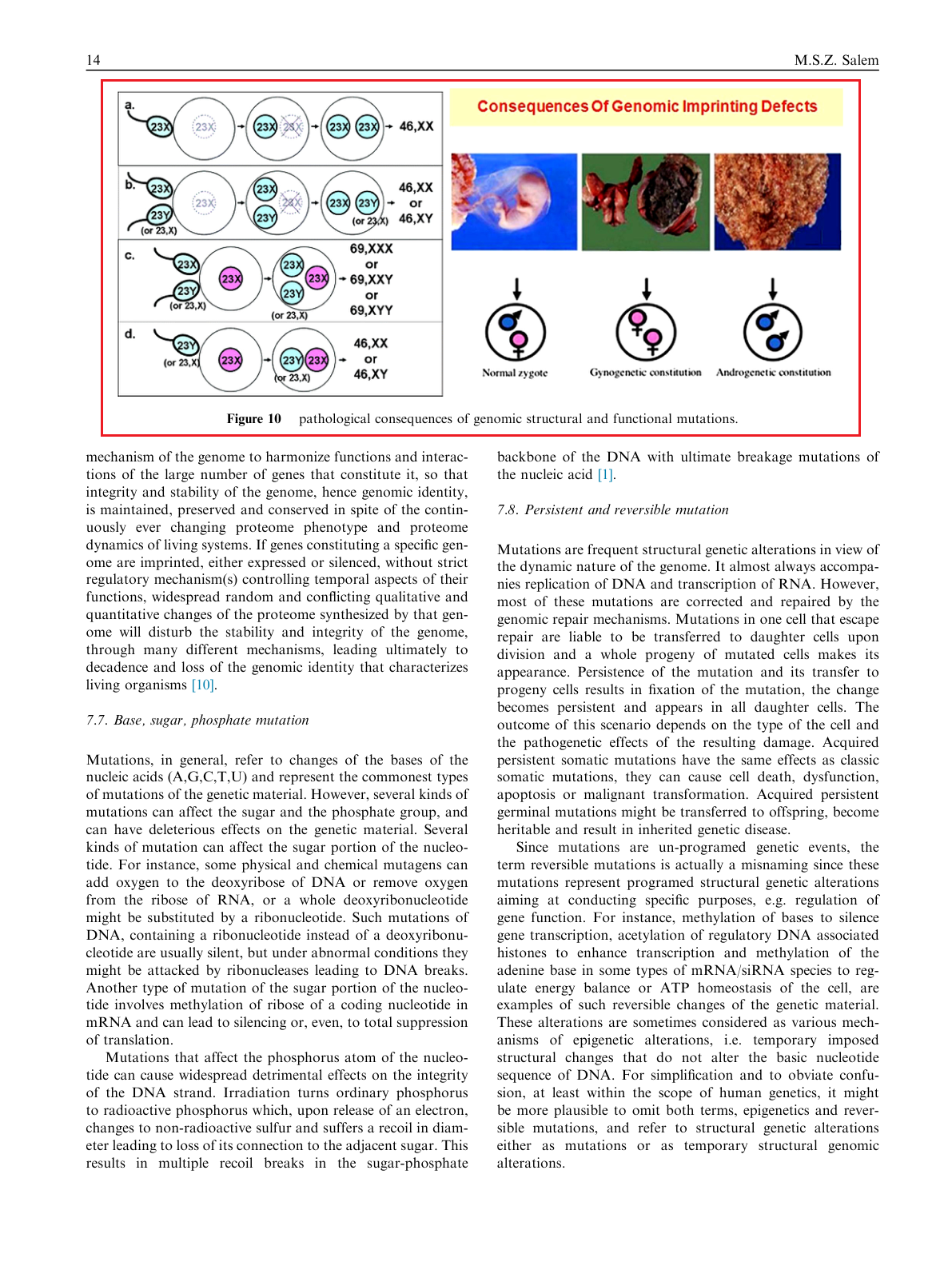}
\par}

\begin{minipage}{\linewidth}
    \textbf{Question:} Which step in Figure 1 maps to the content of Figure 10? \\
    \textbf{Evidence Pages:} [3, 14] \\
    \textbf{Ground Truth:} Deletion/duplication/rearrangement of the genetic material and Genetic diseases.
\end{minipage}

\vspace{2mm} 

\end{tcolorbox}
\label{use case: mldocrag qa example 2}

\end{document}